\pdfoutput=1

\documentclass[11pt]{article}

\usepackage{EMNLP2022}

\usepackage{times}
\usepackage{latexsym}

\usepackage[T1]{fontenc}

\usepackage[utf8]{inputenc}

\usepackage{microtype}

\usepackage{inconsolata}

\usepackage{hyperref}       
\usepackage{url}            
\usepackage{booktabs}       
\usepackage{amsfonts}       
\usepackage{nicefrac}       
\usepackage{xcolor}         

\usepackage{amsmath,amsfonts,bm}
\usepackage{graphicx}
\usepackage{multirow}
\usepackage{makecell}
\usepackage{array}
\usepackage[para]{threeparttable}
\usepackage{booktabs}
\usepackage{tabularx}
\usepackage{xspace}
\usepackage{caption}
\usepackage{subcaption}
\usepackage{bbm}
\usepackage{bm}
\usepackage{enumitem}
\usepackage{algorithm}
\usepackage{algpseudocode}
\usepackage{geometry}
\usepackage{comment}
\usepackage[colorinlistoftodos]{todonotes}
\usepackage{tablefootnote}

\usepackage{amsmath}
\DeclareMathOperator*{\argmax}{arg\,max} 

\usepackage{mathptmx}
\usepackage[11pt]{moresize}

\usepackage{xcolor}

\newcommand{\ours}{GMR\xspace}
\newcommand{\Ours}{Generative Multi-hop Retrieval\xspace}

\newcommand{\pdoc}[1]{d_{\hat{y}_{#1}}}
\newcommand{\gdoc}[1]{d_{y_{#1}}}
\newcommand{\fdoc}[1]{d_{y'_{#1}}}

\title{Generative Multi-hop Retrieval}

\author{Hyunji Lee \quad Sohee Yang \quad Hanseok Oh \quad Minjoon Seo \\
  KAIST AI \\
  \texttt{\{hyunji.amy.lee, sohee.yang, hanseok, minjoon\}@kaist.ac.kr} 
  }

\begin{document}
\maketitle
\begin{abstract}
A common practice for text retrieval is to use an encoder to map the documents and the query to a common vector space and perform a nearest neighbor search (NNS); multi-hop retrieval also often adopts the same paradigm, usually with a modification of iteratively reformulating the query vector so that it can retrieve different documents at each hop. However, such a bi-encoder approach has limitations in multi-hop settings; (1) the reformulated query gets longer as the number of hops increases, which further tightens the embedding bottleneck of the query vector, and (2) it is prone to error propagation. In this paper, we focus on alleviating these limitations in multi-hop settings by formulating the problem in a fully generative way. We propose an encoder-decoder model that performs multi-hop retrieval by simply \emph{generating} the entire text sequences of the retrieval targets, which means the query and the documents interact in the language model's parametric space rather than L2 or inner product space as in the bi-encoder approach. Our approach, \Ours~(\ours), consistently achieves comparable or higher performance than bi-encoder models in five datasets while demonstrating superior GPU memory and storage footprint.\footnote{https://github.com/amy-hyunji/Generative-Multihop-Retrieval}

\end{abstract}
\section{Introduction}
\begin{figure*}[t!]
\small
\centering
\includegraphics[width=0.7\textwidth]{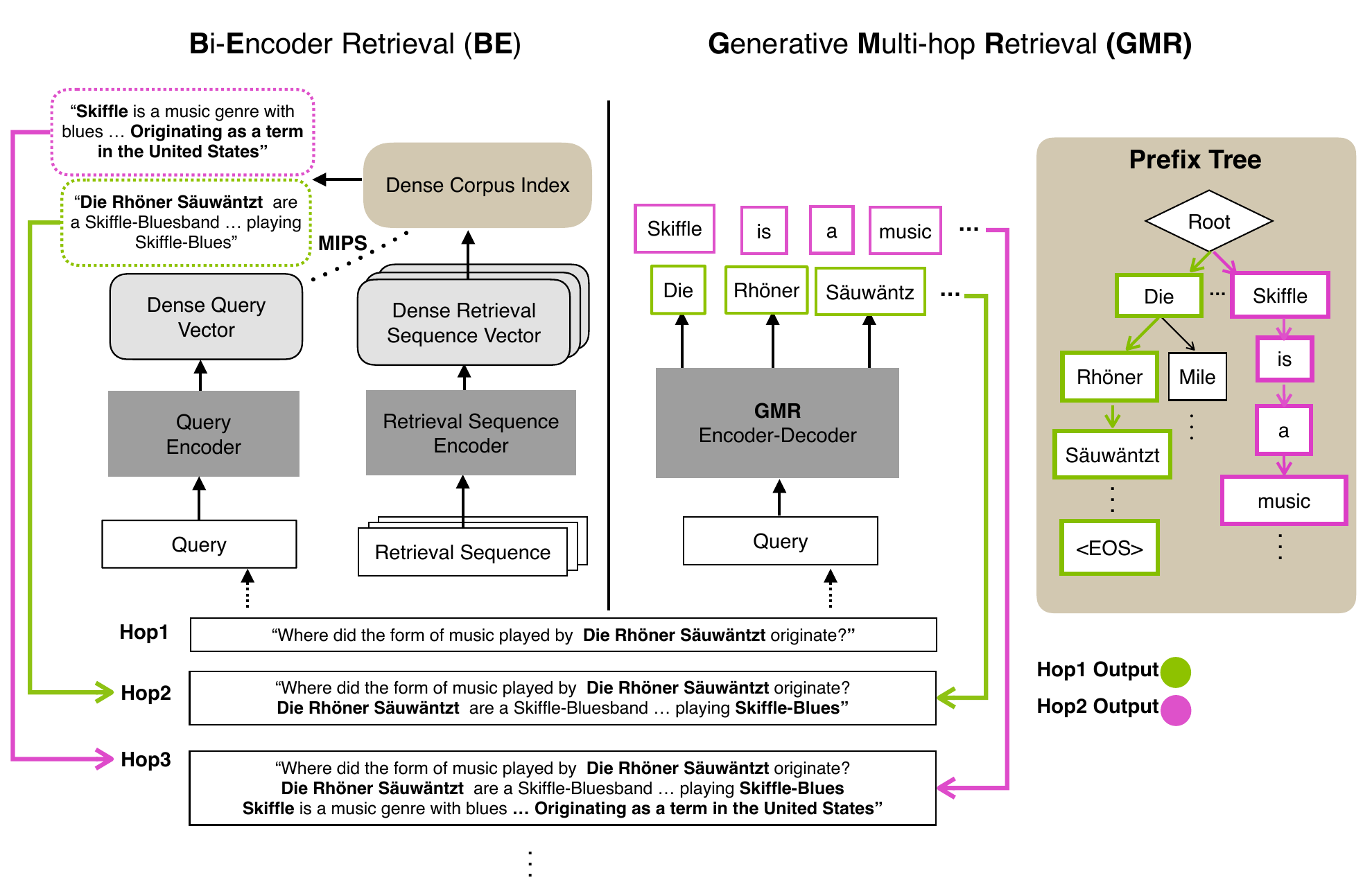}
\caption{\small The figure shows the difference between bi-encoder and \Ours (\ours) for multi-hop retrieval. 
In the first hop retrieval of \ours, to generate the second token, \textit{Rhöner}, it finds the potential next tokens ([Rhöner, Mille]) by searching through the prefix tree with the previously generated tokens. We mask out tokens that are not in the potential next tokens and find the token with the maximum score from unmasked tokens, which in this example is \textit{Säuwäntzt}. Finally, when it retrieves the \textit{<EOS>} token, the generation ends, and the generated output is the retrieval sequence of the query. 
In the second hop retrieval, the concatenation of the query and the previously retrieved sequence is the input query for both approaches.}
\label{fig: overall}
\end{figure*}

\vspace{-0.2em}
Finding the relevant knowledge from a massive collection of information is often formulated as a text retrieval problem.
A large portion of the text retrieval literature focuses on finding the single most relevant paragraph or document (i.e., no hop) to the given query~\citep{karpukhin2020dense, Chen2017ReadingWT}.
When we cannot answer a query with a single document, the task is often formulated as a multi-hop retrieval problem, where one needs to retrieve multiple documents that together provide sufficient evidence to answer the query~\citep{Yang2018HotpotQAAD, Joshi2017TriviaQAAL, Dalvi2021ExplainingAW}.
For example, to answer the question ``Where did the form of music played by Die Rhöner Säuwäntzt originate?'' (Figure~\ref{fig: overall}), we first need to retrieve \textit{the form of music} played by Die Rhöner Säuwäntzt and then \textit{where the form originated from}.

No-hop and multi-hop retrieval tasks are often approached by encoding both the query and retrieval sequences to a common vector space and then finding the sequence whose embedding is closest to the query. This bi-encoder approach for retrieval is often considered as a \textit{de facto} standard; heavy computations such as extracting the dense embeddings of the items in the corpus can be done offline, and one can search over a large number of items with low latency through the nearest neighbor search (NNS) or maximum inner product search (MIPS)~\citep{lewisretrieval, Chen2020ImprovingCQ, Wu2020ScalableZE, Roller2021RecipesFB}. 
While such a bi-encoder approach performs well on many retrieval tasks, it has also shown to suffer from information loss when encoding a long query or document into a fixed-size embedding~\citep{luan2020sparse, izacard2020memory}. 
The problem becomes even more critical in multi-hop retrieval as previously retrieved items are appended to the query while iterating through multiple hops.
The augmented query gets longer as the number of hops increases; therefore, the query embedding gradually becomes incapable of containing the entire information.

In this paper, we argue that a fully generative approach to multi-hop retrieval may be the solution; it overcomes the bottleneck problem by interacting in the whole parametric space of the model trained on the target corpus during the retrieval process, rather than operating on L2 or inner product space as in bi-encoder approach (Figure~\ref{fig: overall}).
We propose \Ours~(\ours), an encoder-decoder model that attempts to memorize the entire target corpus in a generative manner and retrieves the most relevant sequence from the corpus by generating the \textit{entire} sequence with the aid of constrained decoding. We also propose memorization methods (LM and multi-hop memorization) to encourage \ours to memorize the target corpus.
Earlier work in generative retrieval~\citep{Tay2022TransformerMA, cao2021autoregressive, Bevilacqua2022} performs retrieval by generating the entity or the document id that represents the target paragraph or document; \ours instead generates the entire text of the target paragraph, which we believe is more suitable for multi-hop retrieval that requires modeling the interaction between longer queries and more fine-grained text segments.

The main contributions of our paper are that:
\vspace{-0.28em}
\begin{itemize}[noitemsep,leftmargin=1.8em]
    \item We show the limitations of bi-encoder retrieval in multi-hop retrieval tasks: its performance decreases as the number of hops increases and is vulnerable to error propagation.
    \item We show that \Ours (\ours) is robust in solving multi-hop retrieval tasks, performing higher or comparable on five datasets. It is especially strong in multi-hop retrieval settings close to real-world scenarios and datasets with a low unseen rate.
    \item We introduce \emph{multi-hop memorization} which effectively memorizes the target corpus and improves the performance of \ours.
\end{itemize}
\vspace{-0.2em}
Given that generative retrieval shows high performance with high storage efficiency in multi-hop retrieval task compared to the traditional bi-encoder approach, we suggest that generative retrieval has the potential to be a practical alternative for not only the no-hop text retrieval tasks, as shown in \citet{Tay2022TransformerMA, Bevilacqua2022}, but also for multi-hop retrieval tasks as explored in this work.

\section{Related Work}
\vspace{-0.3em}
\paragraph{Multi-hop Retrieval}
Multi-hop retrieval, which answers a query by integrating multiple documents, is often necessary to solve complex queries; it is an active area of research due to its importance.
There has been a line of previous works in multi-hop retrieval with non-textual metadata such as knowledge bases, Wikipedia hyperlinks, or entity linking which leverage such metadata to solve the tasks~\citep{Asai2020Learning, Nie2019RevealingTI, Zhao2020ComplexFQ, Dhingra2020Differentiable}. 
However, they are not expandable to cases where such metadata does not exist. 
Another line of research focuses on expanding the bi-encoder architecture which has shown high performance on no-hop retrieval to multi-hop retrieval~\citep{xiong2021answering,Zhao2021}.  
While such methods have shown good performance, previous studies~\citep{luan2020sparse,izacard2020memory} show that bi-encoder approach suffers from information loss when condensing text into a fixed-size vector.
Since the input text gets longer as the number of hops increases in multi-hop retrieval, it is highly likely for the bi-encoder to fall into the bottleneck problem when the number of hops is large. 
Therefore, to overcome such limitations, it is worth exploring the changes in the fundamental approach; our work suggests that a \textit{generative} method can be an effective alternative to the bi-encoder approach for multi-hop retrieval tasks.
\vspace{-0.2em}
\paragraph{Generative Retrieval}
\citet{cao2021autoregressive} first propose a generative retrieval model, which achieves comparable or higher performances on entity retrieval tasks compared to bi-encoder models.
Moreover, concurrent works DSI~\citep{Tay2022TransformerMA} and SEAL~\citep{Bevilacqua2022} show generative retrieval methods in no-hop retrieval settings. DSI (Differentiable Search Engine) generates structured identifiers of each corpus and shows higher performance than the bi-encoder approach in the NQ dataset.
SEAL (Search Engines with Autoregressive LMs) retrieves an item (paragraph or document) by finding an item containing generated ngram using FM index.
It tests on NQ and KILT benchmarks and shows that the generative retrieval model can even outperform well-designed bi-encoder models such as DPR~\citep{karpukhin2020dense} and GAR~\citep{mao2021generation}\footnote{\footnotesize Note that SEAL and GENRE perform multi-hop QA datasets in no-hop retrieval setting.}.
To see the effectiveness of explicitly generating the entire retrieval sequence, we compare \ours with our re-implementation of DSI\footnote{\footnotesize We re-implement DSI as the source code is not released (Appendix~\ref{app: DSI}).} which we expand to a multi-hop retrieval setting for fair comparison.

\section{\Ours} \label{sec: multi-hop retrieval}
\vspace{-0.3em}
Multi-hop retrieval is a task of retrieving a \textit{set} of sequences (e.g., sentences or paragraphs) from a target corpus $D$ given a query $x$.
It is often approached by iterating through multiple hops where the previously retrieved sequences are appended to the query and form an \emph{augmented query} to model the relationship between the target sequences~\citep{Asai2020Learning, xiong2021answering, Khattab2021BaleenRM, Qi2021AnsweringOQ}.
In this paper, we focus on multi-hop retrieval tasks that resemble a real-world scenario: the oracle number of hops and the correct order of retrieval sequences are not given for each query at the inference time, and the number of oracle hops varies in a wide range.

Canonical text retrieval can be formulated as retrieving a sequence $d_{\hat{y}} = \argmax_{d \in D}P(d|x)$, where $x$ is the query, $d$ is a retrieval sequence in the target corpus $D$, and $\hat{y}$ is the index of retrieved sequence in $D$.
The retrieval is considered successful if $d_{\hat{y}} = d_y$ where $d_y$ is the ground truth target.
On the other hand, multi-hop retrieval aims on finding a \textit{set} of sequences retrieved through $k$ hops, $\mathcal{D}_{\hat{y}} = \{\pdoc{1}, \cdots, \pdoc{k}\}$, given the query.
Here, $\pdoc{i}$ is the sequence retrieved at the $i$-th hop, $\pdoc{i} = \argmax_{d \in D} P(d|x, \pdoc{<i})$, where $x$ is the query and $\pdoc{<i}$ is sequences retrieved at previous hops.
As the canonical text retrieval of a bi-encoder approach is modeled as $\argmax_{d \in D}P(d|x) \propto \argmax_{d \in D} F(d) \cdot G(x)$, which is the inner product between the query vector from an encoder $G$ and the retrieval sequence vector from an encoder $F$, its extension to multi-hop is defined as $P(d|x, \pdoc{<i}) \propto F(d) \cdot G(x, \pdoc{<i})$. 
As the number of hops increases, the augmented query ($x, \pdoc{<i}$) gets longer, and it increases the burden of query encoder $G$ to encode the long augmented query into a fixed-size vector. We investigate that such a burden on the query encoder worsens the \textit{bottleneck problem} of the bi-encoder model and that such models are vulnerable to \textit{error propagation} (Section~\ref{analysis: bi-limitation}).

To alleviate such limitations of the bi-encoder approach in a multi-hop setting, we formulate the problem in a fully \textit{generative} way; the generative approach can interact in the whole parametric space of the model trained on the target corpus during the retrieval process rather than operating only on L2 or inner product space as in bi-encoder approach.
We propose \Ours (\ours), an encoder-decoder model that retrieves the most relevant sequence at each hop from the target corpus by \textit{generating} the sequence using constrained decoding as in the right side of Figure~\ref{fig: overall}. The generation goes over multiple hops to retrieve a set of sequences. In training time, the objective is to maximize:
\vspace{-0.2em}
\begin{gather}
    P((\gdoc{1}, \cdots, \gdoc{n})|x) \propto \prod_{i=1}^{n} P(\gdoc{i}|x,\gdoc{<i}) \\
    = \prod_{i=1}^{n} \prod_{j=1}^{|\gdoc{i}|} P(\gdoc{i}^{(j)} | x, \gdoc{<i}, \gdoc{i}^{(<j)})
\label{eq: new}
\end{gather}
\vspace{-0.2em}
to generate the tokens $\gdoc{i}^{(j)}$ of the ground truth text to retrieve $\gdoc{i}$ at retrieval hops $i = 1, \cdots, n$,\footnote{\footnotesize While multi-hop retrieval targets to retrieve a \textit{set} of sequences, it is oftentimes difficult to train over all possible permutations of the set. Therefore, most previous works assume the order heuristically and train the model.} when the query $x$ and the ground truth target sequences of the previous hops $\gdoc{<i}$ are given as an input to the encoder and all tokens up to the previous step at the current hop ($\gdoc{i}^{(<j)}$) are given as the input to the decoder.
In inference time, \ours decides the sequence to retrieve by $P(d|x, \pdoc{<i}) \propto \prod_{j=1}^{|d|} P(d^{(j)} | x, \pdoc{<i}, d^{(<j)})$, i.e. the probability of generating the token $d^{(j)}$ conditioned on the query $x$, text sequences $\pdoc{<i}$ retrieved until the $i$-th hop, and the tokens $d^{(<j)}$ previously generated at the current hop.
To ensure that the generated sequence is in the corpus, we build a prefix tree and perform constrained decoding with the tree~\citep{cao2021autoregressive}\footnote{\footnotesize The prefix tree is built by aggregating the tokenization result of texts in the corpus. Tokens that create strings that are not a sub-string of any text in the corpus are masked out, and only the next top-k tokens from the unmasked and thus valid set of tokens are passed to the model as the potential next tokens list.}.

Since \ours generates a sequence in a uni-directional way (left to right) during the retrieval process, it cannot know the information at the end of a sequence in advance if the model has not been previously trained to generate the sequence. However, the training set for the target multi-hop retrieval task may not cover all the sequences in the corpus. This may negatively affect the performance, especially when the length of the sequence is long and the training set does not contain enough sequences in the target corpus.
To alleviate the issue, we propose \textit{LM memorization} and \textit{multi-hop memorization}, two corpus memorization methods that aim to store the target corpus in the parameters. By training \ours with the methods, it is able to leverage the memorized information in the parameters on multi-hop retrieval tasks (Appendix~\ref{app: gr_details}). 

\vspace{-0.2em}
\paragraph{LM memorization} \label{sec3: lm-mem}
LM (Language Modeling) memorization is an intermediate task applied before training on a multi-hop retrieval task. 
During LM memorization, \ours is trained on the texts in the corpus using the standard LM objective function: when a corpus $D$ with texts $d$ ($d \in D$) is given, the model learns to maximize the LM probability $P(d) = \prod_{j=1}^{|d|} P(d^{(j)} | d^{(<j)})$ for all $d$ in $D$.
By training the retrieval task on top of the parameters trained on LM memorization, the model is able to be aware of the contents at the end of the sequence it generates beforehand since it has seen the sequence during the LM memorization. 
To make the input of LM memorization similar to that of multi-hop retrieval task, we make the first $m$ (randomly chosen) tokens of the text to generate to serve as the encoder input when maximizing $P(d)$ so that the model is trained to maximize $P(d^{(\ge m)}|d^{(<m)}) = \prod_{j=m}^{|d|} P(d^{(j)} | d^{(<m)}, d^{(\ge m, <j)})$ where $d^{(<m)}$ is the input to the encoder.
\paragraph{Multi-hop Memorization} \label{sec3: multi-mem}
While LM memorization has the benefit that it can be easily applied to GMR on all datasets, one limitation is that it is an \textit{unconditional} generation task (i.e., not depending on a query) while multi-hop retrieval is a conditional generation task where a query is always given. Therefore, we propose an advanced \textit{conditional} variant: multi-hop memorization. 
Multi-hop memorization is a task of maximizing $P({D}_y'|x')$ where $x'$ is a pseudo-multi-hop query generated from a query generation model $Q$\footnote{\footnotesize We use pre-trained T5-large to initialize $Q$.} and ${D}_y' = (\fdoc{1}, \cdots, \fdoc{n})$ is a list of pseudo target sequences.

We perform data augmentation to construct the training data for multi-hop memorization, \{$(x', {D}_y'$\}. First, using the original retrieval dataset \{$(x, {D}_y)$\}, we train a query generator $Q$ to generate query $x$ given the concatenation of ground truth target sequences for the query.
After training $Q$, we sample pseudo-target sequences ${D}_{y'} = (d_{y'_1}, \cdots, d_{y'_k})$ from the target corpus $D$ and generate the corresponding pseudo query $x'$ by feeding ${D}_{y'}$ to $Q$. This sampling-generation step is repeated to create enough set of pairs \{$(x', {D}_y')$\}.

For the method to be beneficial for multi-hop retrieval, we simulate the target distribution of the original dataset $D_y$ when sampling ${D}_{y'}$: each adjoining $d_{y'_i}$ and $d_{y'_{i+1}}$ in ${D}_{y'}$ are relevant.
Therefore, we first find important words or phrases (e.g., entity, subject) $I_d$ for each sequence $d \in D$. Then, for all $d \in D$, we construct ${D}_{y'} = (d_{y'_1}, \cdots, d_{y'_k})$ by first setting $d_{y'_1} = d$. This is an iterative process that $d_{y'_{i+1}}$ is sampled from the set of text sequences $\{d' : |I_{d_{y'_{i}}} \bigcap I_{d'}| > 1$\}, stopped when $|{D}_{y'}|=k$ where $k$ is randomly chosen.

Following the described approach, the constructed dataset \{$(x', {D}_y')$\} becomes similar to the original data. We apply filtering to ensure the quality (Appendix~\ref{app: gr_details}). Since the objective function of multi-hop memorization has the same form with that of multi-hop retrieval, we perform the training in a multi-task manner, using \{$(x', {D}_y')$\} and \{$(x, {D}_y)$\} together as the training data at once, rather than the two-phase training as in LM memorization.

\section{Experimental Setup} \label{sec: setup}
\vspace{-0.3em}
\subsection{Fixed and Dynamic Multi-hop Retrieval} \label{setup: two-setup}
We formulate two settings of multi-hop retrieval tasks: fixed and dynamic multi-hop retrieval settings. 
Our ultimate goal of multi-hop retrieval tasks in the inference step is to retrieve a set of relevant items when given an input query $x$. 
However, since $k$, the oracle number of items in a set, varies depending on $x$ and the task, it is difficult to know $k$ beforehand in a real-world scenario. Therefore, in most cases, $k$ is fixed to a certain number.

Fixed and Dynamic multi-hop retrieval settings differ by whether the retrieval process continues until the maximum retrieval hops $k$ or stops in the middle 
(Appendix \ref{app: fix-dynamic}).
Fixed setting is commonly used in previous multi-hop retrieval tasks, which a model retrieves till the maximum retrieval hop. Whereas dynamic setting is more applicable to solving multi-hop retrieval tasks close to a real-world scenario; rather than iterating until the given maximum number of hops, the model itself predicts when to stop the process by generating the special token (\textit{DONE}) and stops in the middle. 
\vspace{-0.2em}
\subsection{Datasets} \label{setup: dataset}
We use five datasets with various characteristics (Appendix~\ref{app: dataset}). \\
\textbf{HotpotQA}~\citep{Yang2018HotpotQAAD} is an open domain multi-hop question answering dataset, which requires two Wikipedia pages to answer the query. \\
\textbf{Entailment TreeBank (EntailBank)}~ \citep{Dalvi2021ExplainingAW} is a reasoning tree construction task where it forms a tree with a hypothesis as the root node and evidence sentences as leaf nodes. We experiment on Task3: retrieve leaf nodes from the corpus when given a question and an answer as an input. \\
\textbf{StrategyQA}~\citep{Geva2021DidAU} is an open-domain multi-hop question answering dataset where the reasoning steps are implicit in the question. It requires strategies to answer the question. \\
\textbf{Explagraphs-Open (EG-Open)}~\citep{Saha2021ExplaGraphsAE} is a generative and structured commonsense-reasoning task. We reformulate it to open-domain retrieval task (\text{Explagraphs-Open}), which considers a single path (\text{subject-relation-object}) as a retrieval sequence.\\
\textbf{RuleTaker-Open (RT-Open)}~\citep{clark2021transformers} is a synthetic rule-based dataset to measure the model's reasoning ability over rules. We reformulate it to open-domain retrieval task (\text{RuleTaker-Open}) which considers nodes of the graph (sentences) as a retrieval sequence (Appendix~\ref{app: ruletaker-open}).

\vspace{-0.2em}
\subsection{Bi-Encoder Retrieval Models} \label{setup: bi-encoder}
For each dataset, we compare the results with a bi-encoder retrieval model (BE) as a baseline. 
For the HotpotQA dataset, we use MDR
~\citep{xiong2021answering}, a widely used bi-encoder model for the corresponding dataset. For the rest of the datasets, we compare with ST5~\citep{ni2021sentence}.\\
We train and inference BE similar to \ours for both fixed and dynamic settings. In a fixed setting, BE maximize $P(\gdoc{i}|x,\gdoc{<i})$ by concatenating the query $x$ and the retrieval sequences of the previous steps $\gdoc{<i}$ as an input to the query encoder. In a dynamic setting, we add a special token \textit{DONE} to the corpus, and BE is trained to retrieve the special token after the last hop (Appendix \ref{app: bi_details}). \\
\textbf{MDR} is an iterative bi-encoder retrieval model which extends DPR~\citep{karpukhin2020dense} to a multi-hop retrieval. \\
\textbf{ST5} is an encoder-decoder model\footnote{\footnotesize We use ST5-EncDec which extracts sequence embedding by the first output of decoder} that serves as our baseline bi-encoder to compare the performance with \ours using the same number of parameters and architecture, T5~\citep{raffel2020exploring}.

\vspace{-0.2em}

\subsection{Evaluation Metric} \label{setup: metric} 
In the fixed multi-hop retrieval, we evaluate HotpotQA following the MDR evaluation metric\footnote{\href{https://github.com/facebookresearch/multihop_dense_retrieval}{https://github.com/facebookresearch/MDR}}. For the rest, we first calculate the recall rate (R@k) of each query and average over the number of queries~\citep{Dalvi2021ExplainingAW, Saha2021ExplaGraphsAE}.
In the dynamic multi-hop retrieval, since the number of predicted retrieval sequences varies, we measure the F1 score (F1@k) by retrieving a maximum of $k$ sequences.
For RT-Open, we newly define an evaluation metric (Appendix~\ref{app: ruletaker-eval}) that measures the graph construction success rate.

\begin{table*}[t!]
\centering
\fontsize{7.0}{10}\selectfont
\caption
     {\fontsize{6.5}{10}\footnotesize Recall rate (R@5) of fixed setting and F1 score (F1@5, F1@10, F1@20 where each number indicates the maximum retrieval step) of dynamic setting on the test set. We compare results between \ours and ST5 (bi-encoder retrieval) where \ours outperforms ST5 for all four datasets. GMR$_L$ is GMR with LM memorization, and GMR$_M$ is GMR with Multi-hop memorization. The bold text shows the best score of each dataset. Results with * are evaluated by success rate (Appendix~\ref{app: ruletaker-eval}).}
    \begin{tabular}{ccccccccccccccc}
    \toprule
        & \multicolumn{3}{c}{EntailTree} & \multicolumn{4}{c}{StrategyQA} & \multicolumn{4}{c}{EG-Open} & \multicolumn{3}{c}{RT-Open*} \\      
        \cmidrule(lr){2-4} \cmidrule(lr){5-8}
        \cmidrule(lr){9-12} \cmidrule(lr){13-15}
        & ST5 & GMR & GMR$_L$ & ST5 & GMR & GMR$_L$ &GMR$_M$ & ST5 & GMR & GMR$_L$ & GMR$_M$ & ST5 & GMR & GMR$_L$\\
    \midrule
    Fixed R@5 & 31.5 & 53.6 & \textbf{54.3} & 37.4 & 44.9 & 45.5 & \textbf{45.6} & 27.0 & 32.9 & 32.4 & \textbf{34.6} & - &-& -\\
    Dynamic F1@5 & 24.9 & \textbf{48.2} & 47.4 & 38.1 & 41.9 & 42.6 & \textbf{43.1} & 25.0 & 35.5 & 35.7 & \textbf{36.2} & - & - & -\\
    Dynamic F1@10 & 19.4 & \textbf{52.1}& 51.7 & 36.9 & 44.3 & 45.0 & \textbf{45.2}& 24.6 & 40.0& 40.8 & \textbf{42.1} & - & - & - \\
    Dynamic F1@20 & 16.9 & \textbf{52.5}& 52.2 & 36.5 & 46.6 & 47.1 & \textbf{47.9}& 25.4 & 41.5& 41.3 & \textbf{42.6} & 17.0 & 51.0 & \textbf{65.5} \\
    \bottomrule
    \end{tabular}
\label{table: fix-dynamic}
\end{table*}

\section{Experimental Results} \label{sec: results}
\vspace{-0.3em}
In Section \ref{results: results}, we compare the results of \ours and bi-encoder models in fixed and dynamic settings with five different datasets. In Section \ref{results: analysis}, we show the limitations of bi-encoder retrieval models, discuss the effect of unseen rate in \ours, and show \ours's efficiency on storage and inference time. 

\begin{table}[t!]
\centering
\fontsize{7.0}{10}\selectfont
\caption{\fontsize{7.0}{10}\footnotesize Recall rate of HotpotQA official full-wiki dev set. Scores of DPR, MDR- and MDR are from Table 3 of \citet{xiong2021answering}. MDR- indicates a variant of MDR without linked negatives, memory bank, and shared encoder.} 
    \begin{tabular}{l|c c c | c c}
        \toprule
        \textbf{Method} &
        \textbf{DPR} & \textbf{MDR-} & \textbf{MDR} & \textbf{fix-\text{\ours}} & \textbf{fix-\text{GMR$_L$}} \\
        \midrule
            \text{Top-2} & {25.2} & 59.9 & 65.9 & 57.7 & 55.0\\
            \text{Top-10} & 45.4 & 70.6 & 77.5 & 68.8 & 65.3\\
            \text{Top-20} & 52.1 & 73.1 & 80.2 & 73.9 & 71.4 \\
        \bottomrule
    \end{tabular}
    \label{table:hotpot}
\end{table}

\subsection{Results} \label{results: results}

\paragraph{Bi-Encoder (BE) vs. GMR} \label{results: results}
Table~\ref{table: fix-dynamic} shows the overall performance of the bi-encoder baseline (BE) and \ours variants on four datasets (EntailTree, EG-Open, StrategyQA, RT-Open) in fixed and dynamic multi-hop retrieval settings.
We further compare results between our base model (\ours) and \ours with memorization methods: multi-hop memorization (GMR$_M$) and LM memorization (GMR$_L$).
Across all datasets, \ours consistently shows a higher recall rate of top-5 in the fixed setting and a higher F1 score in the dynamic setting than bi-encoder models.
Also, in most cases, both LM memorization and multi-hop memorization methods help improve the performance of \ours (Section~\ref{analysis: unseen}). 
Moreover, the dynamic setting consistently outperforms the fixed setting for both BE and \ours (Appendix~\ref{app: fixed_dynamic_result}), which suggests that the dynamic setting is more adaptable to multi-hop retrieval with a larger number of hops.

Table~\ref{table:hotpot} compares the result between \ours and MDR~\citep{xiong2021answering} on HotpotQA.
While the score of \ours is lower than that of MDR, it is comparable to MDR- (a variant of MDR without linked negative, memory bank, and shared encoder).
One reason why the performance of \ours is similar to MDR-, not MDR, would be that the techniques such as hard negative training or memory bank are crucial for higher performance yet are not applicable to \ours; this suggests an important future direction to close the gap.
Also, since HotpotQA is a fixed to \textit{two}-hop setting, bi-encoder models would suffer less from bottleneck and error propagation problems (Appendix \ref{results: analysis}) compared to the other datasets that require larger numbers of hops.
Results in Table~\ref{table: fix-dynamic} on RT-Open dataset, a task to construct a reasoning graph for the given query in the dynamic setting, suggest that \ours is strong at retrieving sequences interdependent to one another. \ours and $\ours_M$ outperform BE on success rate\footnote{\footnotesize F1 cannot be calculated on RT-Open because the ground truth retrieval sequence is not known at each step.} by 300\% and 385\%, respectively, and construct more complex and diverse reasoning graphs through the retrieval process (Appendix~\ref{app: ruletaker_result}).

While BE needs to create and store a large index of embeddings and often loads it on GPUs for low latency, \ours only needs to create a prefix tree on CPUs, which leads to higher efficiency on offline computation, storage (Appendix~\ref{app: storage}), and GPU memory; \ours shows 69.7\% and 79.5\% decrease of storage and GPU memory, respectively, compared to BE with the same number of parameters (ST5).
During inference time, \ours can be time-inefficient if it has to generate every word in the retrieval target text. In practice, however, one can stop generation as soon as the partially generated text can uniquely identify the target text. By leveraging the optimization, \ours with greedy search is able to achieve a 40\% inference time reduction with respect to ST5 in HotpotQA. Note that without the optimization, \ours is 24.6 times slower than ST5, signifying the importance of early stopping. 

\begin{table}[t!]
    \centering
    \fontsize{6.5}{10}\selectfont
    \caption[Caption for LOF]
      {\fontsize{6.5}{10}\footnotesize Recall rate (R@5) of fixed setting on the test set. We compare results between \ours and \text{DSI*} to show the effectiveness of explicitly generating the entire sequence in a multi-hop retrieval task. \ours outperforms both \text{DSI*} models on all three datasets.}
    \begin{tabular}{lcccc}
        \toprule
        \textbf{Model} & \textbf{EntailTree} & \textbf{StrategyQA} & \textbf{EG-Open}  \\
        \midrule
            \text{atomic-DSI*} & 28.0 & - & 23.4 \\
            \text{naive-DSI*} & 7.7 & - & 8.6\\
            \midrule
            fix-\text{\ours} & 53.6 & 44.9 & 32.9\\
        \bottomrule
    \end{tabular}
    \label{table: DSI}
\end{table}

\paragraph{Importance of Explicit Generation in Multi-hop Retrieval Task}
\ours performs retrieval by explicitly generating the \textit{entire} retrieval sequence using constrained decoding, unlike the previous generative retrieval methods, in order for the retrieval model to better grasp and understand the relationship between the input query and retrieval sequences.
We compare \ours with our implementation of DSI~\citep{Tay2022TransformerMA}, a concurrent work that assigns an id for each document in the corpus and retrieves relevant documents by generating ids. We expand DSI (which experiments only on no-hop settings) to multi-hop settings to retrieve the id of a relevant document and construct an augmented query by adding the text of the retrieved id at the end of the input query as in \ours.
Table~\ref{table: DSI}\footnote{\footnotesize Since DSI is not open-sourced, we reproduced the model ourselves (DSI*). We show NQ results of DSI* in Appendix \ref{app: DSI}. We skip the result of DSI-semantic as we could not reproduce the result. `-' in the table indicates that the model failed on all test cases. We hypothesize it is due to its difficulty in generalization to datasets with a large size corpus. We plan to update the table when the official code of DSI is released.} shows that \ours outperforms DSI on all datasets, implying the benefit of generating the entire sequence in a multi-hop retrieval task.
GENRE~\citep{cao2021autoregressive}, which performs document retrieval by generating the title of the target Wikipedia page with constrained decoding, is not directly applicable to our multi-hop settings since most retrieval sequences in the datasets do not have such titles.

\begin{figure}[t!]
\centering
\includegraphics[width=0.43\textwidth]{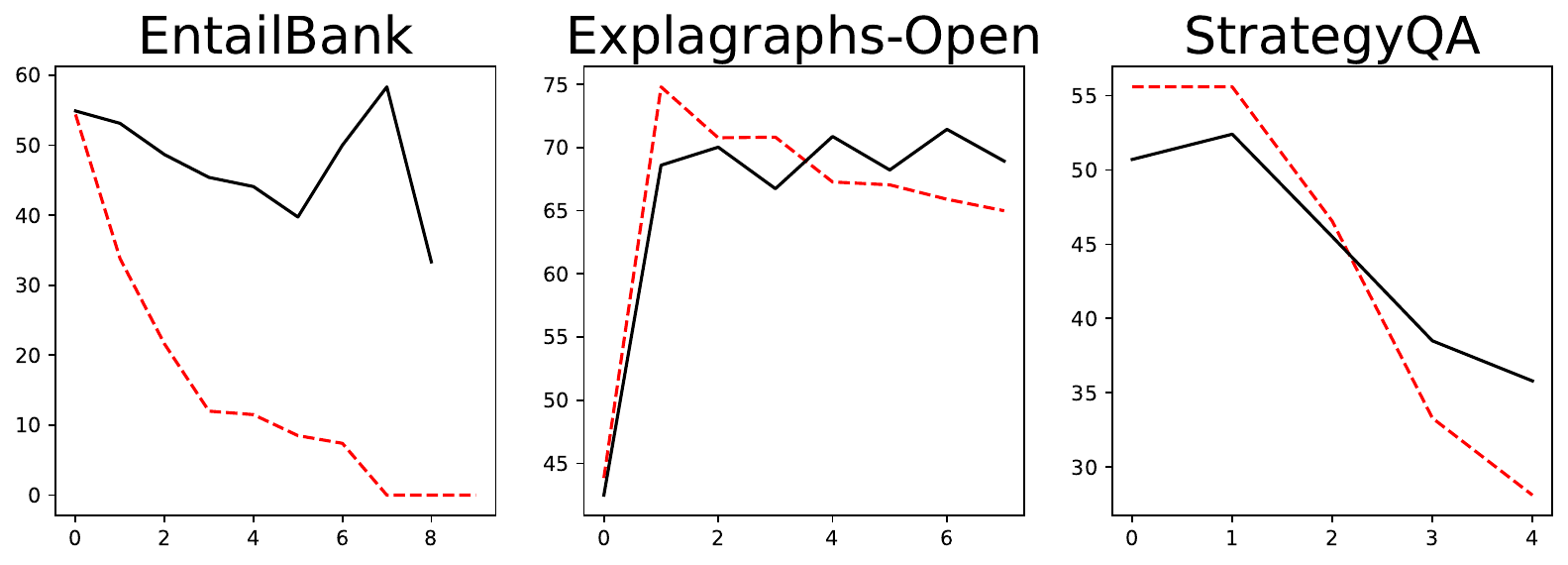}
\caption{\fontsize{6.5}{10}\footnotesize Hop-R@5$_\text{oracle}$ (y-axis) over number of hops (x-axis). The red dotted line and the solid black line show performance of the ST5 (bi-encoder) and \ours, respectively. For all three datasets, ST5 tends to degrade as the number of hops increases, whereas \ours shows relatively consistent performance.}
\label{fig: bi_emb}
\end{figure}

\subsection{Analysis} \label{results: analysis}
\vspace{-0.1em}
\paragraph{Limitation of Bi-Encoder Retrieval Models} \label{analysis: bi-limitation}
We investigate limitations of bi-encoder in multi-hop retrieval and show (1) \textit{bottleneck problem}: performance of bi-encoder consistently decreases as the number of hops increases, and (2) \textit{error propagation}: the bi-encoder approach is more vulnerable to error propagation than generative approach.

For ease of analysis, we compare the performance of the bi-encoder retriever (ST5) and generative retriever (\ours) on three datasets (StrategyQA, EG-Open, EntailBank) under the setting where we assume that a ground truth order of the sequences to retrieve exists. The goal is to retrieve the one gold target sequence $\gdoc{i}$ of each $i$-th hop. The performance is measured as $\text{hop-R@5}_\text{oracle} = \mathbbm{1}{\{\gdoc{i} \in \text{top-5}_{d \in D} P(d|x, \gdoc{<i}) \}}$, where top-5 is a function that returns a set of the five sequences with the largest probabilities. 

\textbf{(1) Bottleneck Problem}
As shown in previous works~\citep{luan2020sparse, izacard2020memory}, bi-encoder approaches have an inherent limitation that their performance degrades proportionally to the size of the embedding. \citet{luan2020sparse} especially shows that the performance decreases more severely as the length of the encoded sequence gets longer. 
We hypothesize that such a limitation would be even more problematic in multi-hop retrieval with a large number of hops; as the previously retrieved sequences are added at the end of the input query, it results in a longer input sequence compared to the canonical retrieval tasks. 

To test the hypothesis, we experiment over ST5 (bi-encoder) and \ours, where the ground truth retrieval targets up to the previous hops are given to seclude the effect of error propagation from the analysis. The result in Figure~\ref{fig: bi_emb} shows that $\text{hop-R@5}_\text{oracle}$ (y-axis) at each hop (x-axis) of the bi-encoder retriever deteriorates more severely than \ours as the number of hops increases.
We could observe that such limitation also occurs in HotpotQA (Appendix~\ref{app: hotpotqa}), where the highest error case of the bi-encoder approach is when it fails to retrieve the second hop correctly although the first hop is correctly retrieved.

It seems like the bi-encoder retriever finds it more difficult to encode the lengthened query into a fixed-size embedding. In contrast, the generative retriever is more robust in modeling the information, possibly because it can mimic the behavior of powerful one-tower cross-encoders (where all tokens in the query and retrieval sequence perform attention to each other), unlike the shallow bottlenecked interaction of bi-encoder. 
We also check that the finding of the previous works still holds in our multi-hop setup: the performance of the bi-encoder retriever monotonically decreases as the embedding size decreases using an additional linear layer at the top of the model (Appendix \ref{app: emb_size}). 

\begin{table}[t!]
\centering
\fontsize{7.0}{10}\selectfont
\caption{\fontsize{6.5}{10}\footnotesize Error propagation rate of a bi-encoder model (BE) and \ours on three datasets: StrategyQA (Str), EG-Open (Exp), and EntailBank (Ent). Details of Minor and Major tasks are in Section~\ref{analysis: bi-limitation}. 
} 
\begin{tabular}{c ccc|ccc}
    \toprule
    & \multicolumn{3}{c}{Minor} & \multicolumn{3}{c}{Major} \\
    \cmidrule(lr){2-4} \cmidrule(lr){5-7}
    & \textbf{Str} & \textbf{Exp} & \textbf{Ent} & \textbf{Str} & \textbf{Exp} & \textbf{Ent} \\
    \midrule
    \text{BE} & 23.6\% & 46.9\% & 14.0\% & 71.2\%  & 91.1\% & 55.1\%\\
    \text{\ours} & 1.7\% & 49.3\% & 11.1\% & 20.7\% & 75.8\% & 39.6\% \\
    \bottomrule
\end{tabular}
\label{table: error}
\end{table}

\textbf{(2) Error Propagation} \label{analysis: error}
We perform experiments to analyze the robustness of the retrievers on error propagation.
We simulate the case where it has retrieved an irrelevant sequence at the previous hop. At $i$-th hop, the retriever is given a query $x$, ground truth retrieval target until the $i-1$-th hop ($\gdoc{<i-1})$, and an irrelevant sequence $d_{\text{error}_{i-1}}$ at the $i-1$-th hop, and we test whether the retriever can still correctly retrieve the ground truth target at the $i$-th hop. We evaluate the robustness of the retrieval model by the error propagation rate ($\text{error propagation rate (\%)} = \left(1-\frac{\text{hop-R@5}_{error}}{\text{hop-R@5}_{oracle}}\right) {\fontsize{7.5}{10}\selectfont \times 100}$), which is the relative drop rate of $\text{hop-R@5}_{error}$ ($\text{hop-R@5}_\text{error} = \mathbbm{1}{\{\gdoc{i} \in  \text{top-5}_{d \in D} P(d|x, \gdoc{<i-1}, d_{\text{error}_{i-1}}) \}}$) from the oracle setup $\text{hop-R@5}_\text{oracle}$.
We experiment over two tasks (major and minor) which differ by how relevant $d_{\text{error}_{i-1}}$ is to the ground truth target at the $i-1$-th hop $d_{y_{i-1}}$: (1) \textit{minor} task is when we find the most relevant sequence excluding the ground truth from the corpus set using BM25 and (2) \textit{major} task is when we randomly sample any sequence from the corpus set.

Results in Table \ref{table: error} indicate that the bi-encoder approach is highly vulnerable to error propagation, where the average error propagation rate is 28.2\% and 72.5\% for minor and major tasks, respectively. 
On the other hand, somewhat surprisingly, \ours is much more robust to error propagation: 20.7\% and 45\%, respectively. 
We hypothesize that such robustness is due to its cross encoding capability over all input tokens (query and retrieval sequences at the previous hops) and its ability to leverage the distribution of sequential text tokens, learned during pretraining, on the retrieval task. 

\vspace{-0.2em}
\paragraph{Effect of Unseen Rate of \ours} \label{analysis: unseen}
The unseen rate indicates the rate of queries in the test set that needs to retrieve sequences never seen during training as the ground truth target. Therefore, datasets with high unseen rates can be considered similar to a zero-shot retrieval setting.
Comparing the relative F1@20 (Table~\ref{table: fix-dynamic}) of \ours to corresponding bi-encoder models in dynamic settings, the improvement is 305.4\% on datasets with low unseen rates (EntailTree, RT-Open), whereas 145.6\% on datasets with high unseen rates (StrategyQA, EG-Open), implying the importance of reducing the unseen rate for \ours. 

We thus train \ours with LM memorization (GMR$_L$) where the model is first trained on target corpus using standard language modeling objective and then finetuned on retrieval task (Section~\ref{sec3: lm-mem}).
Table~\ref{table: fix-dynamic} shows that LM memorization is consistently helpful in StrategyQA and RT-Open, while the gain is inconsistent in other datasets. 
We hypothesize that such inconsistency is because the training objective function of LM memorization is not aligned well with that of multi-hop setting retrieval task; the memorization objective function $P(d) = \prod_{j=1}^{|d|} P(d^{(j)} | d^{(<j)})$ resembles the no-hop retrieval training objective function of maximizing $P(d_y|x)$ rather than that of the multi-hop retrieval, $P((\gdoc{1}, \cdots, \gdoc{n})|x)$, which goes through multiple retrieval hops. Its strength is that it can be easily applied to any dataset, but it does not show consistent improvement on different multi-hop datasets.

We also train \ours with multi-hop memorization (GMR$_M$), where the objective function is similar to that of multi-hop retrieval; we generate pseudo-multi-hop data \{$(x', {D}_y')$\} and use it as additional training data for the retrieval task (Section~\ref{sec3: multi-mem}). The unseen rates of StrategyQA and EG-Open are greatly reduced by applying the method, and GMR$_M$ consistently outperforms both \ours and GMR$_L$: the unseen rates of StrategyQA and EG-Open are reduced by 40.4\% and 60\%, respectively.
While the method is difficult to apply as-is to datasets with low unseen rates due to the filtering process (Appendix \ref{app: pseudo_query}), it is also not necessary as most retrieval sequences in the target corpus are covered by the training set.

\section{Conclusion}
\vspace{-0.3em}
In this paper, we show that the bi-encoder approach has limitations in multi-hop retrieval; the bottleneck problem becomes a more severe problem as the number of hops increases, and is more susceptible to error propagation. 
We present \Ours (\ours), an encoder-decoder model that performs retrieval by \textit{generating} the entire target sequences with the aid of constrained decoding. We show that \ours is more robust on multi-hop retrieval tasks where it achieves higher or comparable performance in five datasets. 
We also introduce two corpus memorization methods, LM memorization and multi-hop memorization, to further improve \ours's performance. Our experimental results demonstrate that in multi-hop retrieval, a generative approach is highly competitive with bi-encoder methods and deserves further explorations in the community. 

\section*{Limitations}
\vspace{-0.2em}
As shown in Table~\ref{table:hotpot}, \ours is still not as good as a well-designed bi-encoder retrieval (MDR) for HotpotQA. We suspect that there are largely two reasons: first, HotpotQA has exactly two hops, whereas \ours seems to be more advantageous when the number of hops is large and dynamic; second, bi-encoder retrieval is a relatively mature research area, whereas generative retrieval is quite new and the community is yet to discover advanced techniques that fully leverage it. 
Early stopping of \ours (Section~\ref{results: results}) helps inference speed but degrades the performance as it is difficult to calculate the total beam score with early stopping; it does not generate till the last token of the target sequence which it also cannot calculate the beam score over all tokens. More research will be needed to achieve both.

\section*{Acknowledgements}
We would like to thank Eunbi Choi, Joel Jang, Miyoung Ko, and Yongrae Jo for helpful discussion. This work was partly supported by Samsung Research grant (2021, Multi-grained Passage Embedding via Cross-to-Bi Encoder Distillation, 40\%) and Institute of Information \& communications Technology Planning \& Evaluation (IITP) grants funded by the Korea government (MSIT) (No.2022-0-00264, Comprehensive Video Understanding and Generation with Knowledge-based Deep Logic Neural Network, 55\%; No.2019-0-00075, Artificial Intelligence Graduate School Program (KAIST), 5\%).

\bibliography{anthology,custom}
\bibliographystyle{acl_natbib}

\appendix

\section{Generative Multi-hop Retrieval} \label{app: gr_details}
\paragraph{LM memorization}
For the path retrieval task (RT-Open, EG-Open), the subject and the relation are given, and the model generates the object of the sentence. 
For paragraph retrieval tasks (HotpotQA, EntailBank, StrategyQA), the first 70\% of the sentence is given as input, and the model generates the rest.
\paragraph{Multi-Hop Memorization}  \label{app: pseudo_query}
For a \textit{conditional} memorization method, we experiment \ours with multi-hop memorization in which we generate pseudo-multi-hop queries $x'$ and train a retriever with not only the original training dataset $\{(x, {D}_{y})\}$, where $x$ is a query and ${D}_{y}$ is a list of target sequences, but also with generated pseudo-datasets $\{(x', {D}_{y'})\}$, where ${D}_{y'}$ is a list of pseudo target sequences, during the retrieval step. To keep the distribution of $\{(x', {D}_{y'})\}$ similar to $\{(x, {D}_{y})\}$, we ensure that elements in the sequence ${D}_{y'}$ are interconnected to one another by constructing the set to have more than one other elements with the same important words or phrases (e.g., entity, subject). 

For StrategyQA, we consider the entities as important words or phrases and extract the entities by NER (Named Entity Recognition)~\footnote{We use NER model provided from huggingface (\href{https://huggingface.co/docs/transformers/main_classes/pipelines}{https://huggingface.co/pipelines})}. We remove sentences that do not contain any entity or contains more than four entities from the target corpus. We found that such removal of sentences with many entities is critical in the performance as sentences with lots of entities tend to be irrelevant to one another although they do have one common entity, which deviates from our original purpose of sampling: to ensure that elements in a sequence ${D}_{y'}$ are interconnected to one another.
For EG-Open, as the items in the target corpus are a path, we consider the subject and the object as important words. 

We add a filtering process for both when sampling a list of pseudo target sequences ${D}_{y'}$ and when generating multiple pseudo-multi-hop queries $x'$. 
When sampling ${D}_{y'}$, we remove sequences that contains all element in a sequence ${D}_{a}$ where ${D}_{a} \in \{{D}_{y}\}$.  
After generating multiple pseudo-multi-hop queries $x'$ with sampled ${D}_{y'}$, we remove queries that do not generate till the end token; the sentence stops the generation in the middle due to its higher beam score compared to when generating till the end. 
We keep the number of items in ${D}_{y'}$ to be within the same range as in ${D}_{y}$. 

In Table~\ref{table: pseudo_query}, we show examples of pseudo-multi-hop data $(\{(x', {D}_{y'})\})$ of StrategyQA and EG-Open. By adding pseudo-multi-hop data to the original training dataset, we could reduce the unseen rate of StrategyQA and EG-Open by 40.4\% and 60\%, respectively, and increase the performance on both datasets. While the method is difficult to apply to datasets with low unseen rates due to the filtering process of removal when a set of target sequence contains a set in the original retrieval dataset as a subset, it is also not necessary as most retrieval sequences in the target corpus are covered by the training set. We leave the method of generating effective pseudo-multi hop data for datasets with low unseen rate as future work.

\begin{table*}[t!]
\centering
\fontsize{7.5}{10}\selectfont
\caption{\fontsize{7.5}{10}\footnotesize Examples of pseudo-multi-hop data. The top two examples are pseudo-multi-hop data from StrategyQA and the bottom two examples are from EG-Open dataset.} 
\begin{tabular}{ m{6cm} m{9cm}} 
    \toprule
    \textbf{Input} & \textbf{Output} \\
    \midrule
        \multirow{6}{=}{\\[4em] Did Mary, Queen of Scots know Jesus?} &
        Output 1 \\
        \cmidrule(lr){2-2}
        & Mary, Queen of Scots was Queen of Scotland in the 1500s\\
        \cmidrule(lr){2-2}
        & Output 2\\
        \cmidrule(lr){2-2}
        & According to the Gospel of Matthew, Joseph and Mary resettled in Nazareth after returning from the flight from Bethlehem to Egypt\\
        \cmidrule(lr){2-2}
        & Output 3\\
        \cmidrule(lr){2-2}
        & She accompanied Joseph to Bethlehem, where Jesus was born\\
    \midrule
        \multirow{6}{=}{\\[3.0em]Is the iPhone still the most popular computer brand in the world?} &
        Output1 \\
        \cmidrule(lr){2-2}
        & All generations of the iPhone use Apple's iOS mobile operating system software\\
        \cmidrule(lr){2-2}
        & Output2\\
        \cmidrule(lr){2-2}
        & "upset" about the price drop, Apple gave store credit to early adopters\\
        \cmidrule(lr){2-2}
        & Output3\\
        \cmidrule(lr){2-2}
        &Apple stores stock only Mac brand computers\\
    \midrule
        \multirow{6}{=}{\\[3.0em]belief: Everyone is abusive. / argument: Some people are just harmful to others.} &
        Output1 \\
        \cmidrule(lr){2-2}
        & everyone; synonym of; people\\
        \cmidrule(lr){2-2}
        & Output2\\
        \cmidrule(lr){2-2}
        & people; not has property; abusive\\
        \cmidrule(lr){2-2}
        & Output3\\
        \cmidrule(lr){2-2}
        &abusive; synonym of; harmful\\
    \midrule
        \multirow{6}{=}{\\[3.0em]belief: Social media is negative. / argument: Social media keeps us connected.} &
        Output1 \\
        \cmidrule(lr){2-2}
        & social media; causes; connected\\
        \cmidrule(lr){2-2}
        & Output2\\
        \cmidrule(lr){2-2}
        & connected; capable of; causing content\\
        \cmidrule(lr){2-2}
        & Output3\\
        \cmidrule(lr){2-2}
        &causing content; not capable of; negative\\
    \bottomrule
\end{tabular}
\label{table: pseudo_query}
\end{table*}

\section{Experimental Setup}
\subsection{Fixed and Dynamic Multi-hop Retrieval} \label{app: fix-dynamic}

\begin{algorithm}[t!]
\centering
\fontsize{7.5}{10}\footnotesize
\caption{\fontsize{7.5}{10}\footnotesize Inference step of Fixed Conditional Retrieval}
\begin{algorithmic}
    \Require{trained retriever $R$, fixed number of iteration step $k$, input query $q$, and target corpus $\{d_i\}$}
    \\
    \State $Y'$ := An empty set to store all retrieved sequences
    \State $s$ := a number of iteration step with an initial value of 1
    \\
    \While{$s \leq k$}
        \State $y' = R(x, Y')$
        \State $Y'$.add($y'$)
    \EndWhile
    \\
    \State return $Y'$
\end{algorithmic}
\label{alg: fixed}
\end{algorithm}

\begin{algorithm}[t!]
\centering
\fontsize{7.5}{10}\footnotesize
\caption{\fontsize{7.5}{10}\footnotesize Inference step of Dynamic Conditional Retrieval}
\begin{algorithmic}
    \Require{trained retriever $R$, fixed number of iteration step $k$, input query $q$, and target corpus $\{d_i\} \cup \{\emptyset\}$}
    \\
    \State $Y'$ := An empty set to store all retrieved sequences
    \State $s$ := number of iteration step which initial value is 1
    \\
    \While{$s \leq k$}
        \State $y' = R(x, Y')$
        \\
        \State {\color{red} {// If retriever retrieves null element, stop the iteration and fill the set with the null element}}
        \If{$y' == \emptyset$}
            \State Fill $Y'$ with $\emptyset$ so that $len(Y') == k$
            break
        \Else
            \State $Y'$.add($y'$)
        \EndIf
        \\
    \EndWhile
    \\
    \State return $Y'$
\end{algorithmic}
\label{alg: dynamic}
\end{algorithm}

We formulate two settings of multi-hop retrieval: fixed and dynamic multi-hop retrieval settings. 
In the inference step of the multi-hop retrieval, we eventually aim on retrieving a set of items by retrieving for $k$ hops when given an initial input query $x$. 
However, since $k$ varies depending on $x$ and task, it is difficult to know the exact number of $k$ beforehand in a real-world scenario. 
Therefore due to this limitation of real-world setting, we fix the number of retrieval step $k$ for all datasets and queries, which is $k=5$ in this paper.\footnote{We fix k to 5 since it is near the average for all datasets.} 
Fixed conditional retrieval task is an expansion of canonical text retrieval task and dynamic conditional retrieval task is more likely on solving multi-hop retrieval task by stopping the retrieval process before filling all $k$.
In fixed multi-step, the number of items to retrieve is given as an oracle, whereas in dynamic multi-step, the model also needs to determine when to stop retrieving the next item.  

\paragraph{Fixed Multi-hop Retrieval Setting}
\textit{Fixed Multi-hop Retrieval Setting} is a basic setup for multi-hop retrieval tasks, where we assume the given $k$ as the oracle number of hops and iterate till the maximum retrieval hop $k$.
In training process, when given a query $x$ with 3 elements in an oracle set $\mathcal{D}_{y} = \{\gdoc{1}, \gdoc{2}, \gdoc{3}\}$, the training examples for the query is $\{(x, \gdoc{1}), (x; \gdoc{1}, \gdoc{2}), (x; \gdoc{1}; \gdoc{2}, \gdoc{3})\}$ where ; is a string concatenation operator and each element is (input, ground truth output).
In the inference process, when given a trained retriever and a target corpus, it iterates $k$ times to find a set of retrieval sequences related to a given query as in Algorithm \ref{alg: fixed}.

\paragraph{Dynamic Multi-hop Retrieval Setting}
\textit{Dynamic Multi-hop Retrieval Setting} is a setting where the retriever has to predict the correct number of oracle hops $k$ (when to stop retrieving). It differs from the fixed multi-hop retrieval setting, which assumes that the oracle number of hops for each input query is given and fixed. 
It can be considered efficient compared to the commonly used fixed multi-hop setting because (1) in real-world scenarios, we may not know the exact number of texts to retrieve in advance and the number differs by the query; (2) the model has to retrieve till the maximum hop even if it has already retrieved all the relevant sequences in fixed multi-hop setting, which would not only cause unnecessary extra time and computation but also harm the downstream tasks performance~\citep{Qi2021AnsweringOQ} by providing hard negative (false positive) sequences which are difficult to distinguish to the downstream module. 
The additional ability to detect the stopping point of the retrieval hop (dynamic multi-hop retrieval) can resolve the fixed multi-hop retrieval issues above.

For the setting, we additionally add a null element ($\emptyset$)\footnote{We add special token DONE to the model and the corpus which is the null element during training and inference.} to the given target corpus in fixed multi-hop setting $D$; in this setup, a retrieved sequence is an item in $D \cup \{\emptyset\}$. 
In the training process, for a query, we add one extra step at the end of each end step from the fixed multi-hop retrieval training set: given all oracle sequences as conditions, retrieve null element $\emptyset$. 
For example, when given a query $x$ with 3 elements in an oracle set  $\mathcal{D}_{y} = \{\gdoc{1}, \gdoc{2}, \gdoc{3}\}$, the training examples for the query is $\{(x, \gdoc{1}), (x; \gdoc{1}, \gdoc{2}), (x; \gdoc{1}; \gdoc{2}, \gdoc{3}),$ $(x;\gdoc{1};\gdoc{2};\gdoc{3}, \emptyset)\}$ where ; is a string concatenation operator and each element is (input, ground truth output).
In the inference process, when given a trained retriever and a target corpus, as in fixed multi-hop retrieval, it iterates $k$ times to find a set of retrieval sequences related to a given query. However, when the retriever retrieves the null element, it ends the iteration as in Algorithm \ref{alg: dynamic}.

\subsection{Datasets} \label{app: dataset}
\subsubsection{Datasets Details}
Table~\ref{table:dataset}\footnote{For RT-Open unseen rate, we calculate it with prediction result since there are no gold retrieval sequences.} shows the overview of the five datasets. HotpotQA can be download in \href{https://hotpotqa.github.io/}{https://hotpotqa.github.io/}, and the rest of the dataset can be download in \href{https://allenai.org/data}{https://allenai.org/data}. Note that all datasets are in English. 

\begin{table}[t!]
    \centering
    \fontsize{7.5}{10}\selectfont
    \vspace{1.0em}
    \caption{\fontsize{7.5}{10} Overview of the five datasets. 
    \textbf{Seq Len} column shows the average number of retrieval sequence tokens for each retrieval sequence in given target corpus. 
    \textbf{Unseen Rate} column shows the rate of test queries consisting of only the retrieval sequences unseen during the training process.} 
     \vspace{1.0em}
    \begin{tabular}{ c c c c c r}
        \toprule
        \textbf{Dataset} &
        \textbf{Corpus (MB)} & \textbf{Seq Len} &  \textbf{Unseen Rate}\\
        \midrule
            HotpotQA & 1,595 & 78.6  & 18.9\%\\
            EntailBank & 0.7 & 12.5   & 2.7\%\\
            StratgyQA & 7.0 & 13.1  & 98.2\%\\
            EG-Open & 0.5  & 9.6  & 95.5\%\\
            RT-Open & 0.7 & 13.1  & 0.0\% \\
        \bottomrule
    \end{tabular}
    \label{table:dataset}
\end{table}

\paragraph{HotpotQA}
\citet{Yang2018HotpotQAAD} propose an open domain multi-hop question answering dataset, which requires aggregating multiple Wikipedia passages through logical reasoning or sequential processing. The number of retrieval sequences is fixed to two. 
HotpotQA consists of two types of questions: comparison and bridge. Comparison questions, a rationale/evidence type of multi-hop dataset, do not necessitate iterative retrieval since the two entities can be retrieved by the query itself. However, bridge questions consist of evidence in the reasoning chain from where it has to retrieve the second step based on the first one.
We use the official Wikipedia dump provided by \citet{Yang2018HotpotQAAD}, use 2\% of the official train dataset as a dev set, and report the scores on the official dev set.

\paragraph{Entailment TreeBank (EntailBank)}
\citet{Dalvi2021ExplainingAW} propose a reasoning tree construction task where it forms a tree with a hypothesis as the root node and evidence sentences are leaf nodes.
The dataset has three settings, and among them, we experiment on Task3, an open setting. Task3 consists of two steps; the first is to select a leaf node from the corpus set when given a question and an answer, and the second is to construct a reasoning tree through the selected leaf node. We perform the first step, the leaf node retrieval. Since the leaf node and the root node are not directly connected, there is a less tight connection between the input query and gold outputs than in other datasets. 
We experiment on the first step of Task3 (leaf node retrieval). 
As in the paper, we use both EntailBank and WorldTreeV2 \citep{xie2020worldtree} datasets when training a retrieval model.
We compare the results with ST5 since there is no released bi-encoder model, and as in the paper, we use both EntailBank and WorldTreeV2 \citep{xie2020worldtree} datasets when training a retrieval model.

\paragraph{StrategyQA}
\citet{Geva2021DidAU} propose a multi-hop open-domain question answering dataset where the reasoning steps are implicit in the question and need some strategy to answer the question. When given a question, the model retrieves the evidence sentences from the corpus. 
Since only the train dataset contains evidence annotation, we split it into 75/5/20 (\%) and used it as a train/val/test set, respectively. 
Also, based on the given corpus, we split the given paragraph-level corpus to sentence level using NLTK \citep{bird2009natural} to match the granularity of the evidence and add the annotated evidence sentences to the corpus.

\paragraph{RuleTaker-Open (RT-Open)}
\citet{clark2021transformers} propose a synthetic rule-based dataset to measure the model's reasoning ability over the rules expressed in natural language. Based on the released dataset, we create a new task, \text{RuleTaker-Open}, to make the task close to a real-world setting. Given a query, the model retrieves nodes of the graph, which is a sentence from the corpus, and the nodes are connected in order to construct a graph.
Details of the construction method are described in Appendix \ref{app: ruletaker-open}.

\paragraph{Explagraphs-Open (EG-Open)}
\citet{Saha2021ExplaGraphsAE} propose a generative and structured commonsense-reasoning task. When given a belief and an argument, a model predicts whether the argument supports or counters the belief and generates (retrieves) a reasoning graph to explain the prediction. While the original dataset needs generation on constructing the reasoning graph, which is limited to generative models only, we expand the task to an open-domain retrieval setting to compare with the bi-encoder models by constructing the corpus and name it \text{Explagraphs-Open}. We consider a single path (\textit{subject-relation-object}) as a retrieval unit and construct the corpus by dumping all the possible paths provided from the dataset.

\subsubsection{Datasets Examples}
Examples of each dataset (input and output forms) are in Table \ref{table:Data Example}.

\newcolumntype{P}[1]{>{\centering\arraybackslash}p{#1}}
\newcolumntype{M}[1]{>{\centering\arraybackslash}m{#1}}

\begin{table*}[t!]
\centering
\fontsize{7.5}{10}\selectfont
 \vspace{1.0em}
\caption{Dataset examples} 
 \vspace{1.0em}
\begin{tabular}{ c m{5.7cm} m{5.7cm}}
    \toprule
    \textbf{Task} & \textbf{Input} & \textbf{Output} \\
    \midrule
        \multirow{4}{*}{\text{\makecell{\\[4em]Paragraph Retrieval \\(HotpotQA)}}} &
        Step 1 Input (a query) & Step 1 output (evidence passage)\\
        \cmidrule(lr){2-3}
        & <QUESTION> The Oberoi family is part of a hotel company that has a head office in what city? </QUESTION> & <TITLE> Oberoi family </TITLE> The Oberoi family is an Indian family that is famous for its involvement in hotels, namely through The Oberoi Group.\\
        \cmidrule(lr){2-3}
        & Step 2 Input (a query with previous output) & Step 2 Output (evidence passage)\\
        \cmidrule(lr){2-3}
        & <QUESTION> The Oberoi family is part of a hotel company that has a head office in what city? </QUESTION> <EVIDENCE> <TITLE> Oberoi family </TITLE> The Oberoi family is an Indian family that is famous for its involvement in hotels, namely through The Oberoi Group. </EVIDENCE> & <TITLE> The Oberoi Group </TITLE> The Oberoi Group is a hotel company with its head office in Delhi. Founded in 1934, the company owns and/or operates 30+ luxury hotels and two river cruise ships in six countries, primarily under its Oberoi Hotels \& Resorts and Trident Hotels brands.\\
    \midrule
        \multirow{6}{*}{\text{\makecell{\\[8em]Sentence Retrieval \\(EntailmentBank,\\ StrategyQA)}}} &
        Step 1 Input (a query) & Step 1 output (evidence sentence)\\
        \cmidrule(lr){2-3}
        & <QUESTION> Does a dentist treat Bluetooth problems? </QUESTION> & A dentist is a surgeon who specializes in dentistry, the diagnosis, prevention, and treatment of diseases and conditions of the oral cavity\\
        \cmidrule(lr){2-3}
        & Step 2 Input (a query + Step 1 Output) & Step 2 Output (evidence sentence)\\
        \cmidrule(lr){2-3}
        & <QUESTION> Does a dentist treat Bluetooth problems? </QUESTION> <EVIDENCE> A dentist is a surgeon who specializes in dentistry, the diagnosis, prevention, and treatment of diseases and conditions of the oral cavity </EVIDENCE> & Technological problems are typically handled by IT professionals\\
        \cmidrule(lr){2-3}
        & Step 3 Input (a query + Step 1 \& Step 2 Output) & Step 3 Output (evidence sentence)\\
        \cmidrule(lr){2-3}
        & <QUESTION> Does a dentist treat Bluetooth problems? </QUESTION> <EVIDENCE> A dentist is a surgeon who specializes in dentistry, the diagnosis, prevention, and treatment of diseases and conditions of the oral cavity </EVIDENCE> <EVIDENCE> Technological problems are typically handled by IT professionals </EVIDENCE> & Bluetooth is not a physical entity\\
    \midrule
        \multirow{4}{*}{\text{\makecell{\\[7em]Reasoning Path Retrieval \\(RuleTakers,\\ Explagraphs)}}} &
        Step 1 Input (a query) & Step 1 output (evidence sentence)\\
        \cmidrule(lr){2-3}
        & <QUESTION> belif: marriage is the best for a family unit. argument: Marriage is a predictor of health and happiness. </QUESTION> & marriage; created by; love\\
        \cmidrule(lr){2-3}
        & Step 2 Input (a query + Step 1 Output) & Step 2 Output (evidence sentence)\\
        \cmidrule(lr){2-3}
        & <QUESTION> belif: marriage is the best for a family unit. argument: Marriage is a predictor of health and happiness. </QUESTION> <EVIDENCE> marriage; created by; love </EVIDENCE> & love; causes; health and happiness \\
        \cmidrule(lr){2-3}
        & Step 3 Input (a query + Step 1 \& Step 2 Output) & Step 3 Output (evidence sentence)\\
        \cmidrule(lr){2-3}
        & <QUESTION> belif: marriage is the best for a family unit. argument: Marriage is a predictor of health and happiness. </QUESTION> <EVIDENCE> marriage; created by; love </EVIDENCE> <EVIDENCE> love; causes; health and happiness </EVIDENCE> & health and happiness; used for; family unit\\
    \bottomrule
\end{tabular}
\label{table:Data Example}
\end{table*}

\subsection{Details of RuleTaker-Open (RT-Open)}  \label{app: ruletaker-open}

\begin{table}[t!]
    \centering
    \fontsize{7.5}{10}\selectfont
    \caption{\label{app: ruletaker-error} Error rate for each error type in RT-Open. Results are from 200 test sets.}
    \begin{tabular}{lcc}
    \toprule 
    \textbf{Error Rate (\%)} & \textbf{\ours} & \textbf{ST5} \\ 
    \midrule
    Node Num Error & 0.5 &  5\\
    Start Node Error & 9.5 & 0 \\
    End Node Error & 20 & 28\\
    Missing Edge Error& 19 & 50 \\
    Success & 51 & 17 \\
    \bottomrule
    \end{tabular}
\end{table}

\newcommand{\factorial}{\ensuremath{\mbox{\sc Factorial}}}
\begin{algorithm}[t!]
\caption{\fontsize{7.5}{10}\footnotesize Finding the missing edge}\label{missing_edge}
\centering
\fontsize{7.5}{10}\selectfont
\begin{algorithmic}
    \Require{Input Corpus $P$}
    \State $T$ := An empty list to append or remove facts from $P$\\
    \ForAll {sentence $s$ $\in$ $P$}
        \If{$s$ is a rule} 
            \State{divide $s$ to assumptions $A$ and result $r$}
            \ForAll{assumption $a$ $\in$ $A$}
                \If{$a$ in $T$}
                    \State{$T$.remove($a$)}
                \Else
                    \State{return False} \Comment{Missing edge}
                \EndIf
            \EndFor
            \State{$T$.append($r$)}
        \Else
            \State{$T$.append($s$)}
        \EndIf
    \EndFor\\
    \If{$T$ is empty}
        \State{return True} \Comment{No missing edge}
    \Else
        \State{return False} \Comment{Missing edge}
    \EndIf
\end{algorithmic}
\end{algorithm}

RuleTaker dataset is a synthetic rule-based dataset used to measure the model's ability on reasoning over rules~\citep{clark2021transformers, tafjord2021proofwriter, saha2020prover}. Given a small corpus of textual facts and rules, the model has to answer the question, retrieve, and construct the graph-structured proofs. As in \citet{tafjord2021proofwriter}, we use the maximum depth dataset \textit{D5} for training. \\
To evaluate the model performance in the open-setting (i.e., Task3 in \citet{Dalvi2021ExplainingAW}), we newly construct a large corpus and divide the train/dev/test dataset by the unique query set from the original D5 dataset. \\
\paragraph{Dataset Construction}
We dump all the facts and rules from the original D5 train/dev/test datasets to construct the corpus and collect 1621 unique queries, which we split into 1300/121/200. We remove cases with \textsc{NAF} and \textsc{FAIL} cases for rule-based evaluation, remove graphs with less than two nodes to ensure that the fact from the corpus itself could not be the proof, and remove graphs with more than ten nodes to fit in the maximum length of T5 model. Also, we added \textit{DONE} at the end of graph construction for dynamic stopping.\\
\paragraph{Evaluation Metric}  \label{app: ruletaker-eval}
In RT-Open, there are various possible answer graphs for a query, unlike the previous RuleTaker dataset. 
Therefore, to check whether the prediction graph is correct, a new evaluation metric is necessary.
Since each textual sentence can be divided into a simple format, \textit{subject-relation-object}, when considering the constructed method \citep{clark2021transformers}, we evaluate the result by a new rule-based method. \\
We check whether the constructed graph is well-constructed by four steps.
\begin{itemize}[noitemsep]
    \item \textit{Node Num Error}: The number of evidence should be larger than 2. 
    \item \textit{Start Node Error}: First word (\textit{subject}) should be the same.
    \item \textit{End Node Error}: Last word (\textit{object}) should be the same.
    \item \textit{Missing Edge Error}: There should be no missing edge.
\end{itemize}
Table \ref{app: ruletaker-error} shows the rate on each constraint for both the bi-encoder model and \ours. Each error in the table corresponds to the item on top with the same name.\\ 
\textit{Missing Edge Error} is evaluated by Algorithm \ref{missing_edge}; when given a prediction graph ($P$), we divide the sentences into rules and facts and check for the missing edge in the prediction order. When the algorithm returns \textit{True}, the graph is considered to have no missing edge.

\subsection{Bi-Encoder Retrieval Models} \label{app: bi_details}
We use \text{ST5} model~\citep{ni2021sentence} as the architecture of the bi-encoder baseline to compare the performance with \ours using the same number of parameters. The input text is fed into T5-encoder, and the first decoder output of the T5-decoder is taken as the sentence embedding. We follow the implementation details in \citet{ni2021sentence} except for two settings: (1) as in \citet{karpukhin2020dense}, we use the inner product instead of cosine similarity when calculating the similarity since inner produce shows a higher recall rate than cosine similarity for overall dataset (2) we change the hyperparameters for a fair comparison with \ours. 

\subsection{Details} 
\label{app: hyperparameter}
We train both ST5 and \ours using pre-trained T5-large checkpoint (770 million parameters) from \citet{Wolf2020TransformersSN} as the initial checkpoint. 
We use the same hyperparameter setting when training \ours and ST5 model for a fair comparison. 
We observe that hyperparameter change does not change the tendency of results after experimenting over a combination of settings used in previous models~\citep{karpukhin2020dense, ni2021sentence, raffel2020exploring}. 
Also, we use different hyperparameters for different tasks: retrieval corpus memorization and retrieval. For all experiments, we use 8 32GB V100 GPUs.
In retrieval task, training a epoch of HotpotQA take 1.5 hours and for the rest it take less than 0.5 hours.

\paragraph{LM Memorization} 
The LM memorization step aims to show \ours a corpus it will retrieve and sa it implicitly before the retrieval step. We keep the learning rate to 1e-5, which is relatively low than the retrieval step, to maintain the linguistic ability the model learned during pretraining~\citep{jang2022towards}. We train the model from T5 pre-trained checkpoint for every dataset using Adafactor with a constant learning rate of 1e-5 with batch size 240 till the maximum of 3 epochs.

Increasing the LM memorization epoch does not always lead to higher performance. This is because as the model is trained on a new dataset, catastrophic forgetting of previously learned parts occurs~\citep{Kirkpatrick2017OvercomingCF}, and in this case, the linguistic ability of the model learned during the pretraining step. To prevent the following process from occurring, we follow \citet{jang2022towards} and reduce the learning rate to 1e-5 and 
use checkpoint of epoch 3 as the initial checkpoint for all retrieval tasks.

\paragraph{Multi-Hop Memorization}
We train a model that generates a pseudo-multi-hop query for multi-hop memorization when given a set of retrieval sequences. 
We dump all retrieval datasets to train such a model and concatenate all retrieval sequences as a long sequence as an input and the corresponding query as an output.
Generated pseudo-multi-hop queries after the filtering process are 11k and 1.9K for StrategyQA and EG-Open, respectively. 
We set the configuration the same as in \textit{Retrieval Step}.

\paragraph{Retrieval Step}
The retrieval step aims to retrieve the gold item from a large-scale corpus. For \ours with LM memorization ($\ours_L$), we use the checkpoint from LM-memorization as the initial checkpoint, and for the rest of the models (ST5, \ours, \ours with multi-hop memorization ($\ours_M$)), we use the T5 pre-trained checkpoint as the initial checkpoint.
For $\ours_M$, we train the model using both the training dataset and generated dataset from the T5 pre-trained checkpoint.
For both ST5 and \ours (including $\ours_L$, $\ours_M$), we train using Adafactor with a learning rate 1e-4 with a linear warm-up for the first 10\% of training and then linear decay with batch size 120 till a maximum of 30 epochs. 

\section{Experimental Results}
\subsection{Results}

\begin{table}[t!]
    \centering
    \fontsize{7.5}{11}\selectfont
    \caption{\fontsize{7.5}{10}\footnotesize Retrieval sequence F1 score of model trained on fixed multi-hop retrieval setting (*-fix) and dynamic multi-hop retrieval setting (*-dyn) on test set. 
    F1@k is a retrieval sequence F1 score with maximum retrieval step of k. Models trained on dynamic setting show consistently higher F1 score compared to those trained in fixed setting.}
    \begin{tabular}{cl@{\hskip 0.1in} c c c}
        \toprule
        \textbf{Dataset} & \textbf{Model} & 
        \textbf{F1@5} & \textbf{F1@10} & \textbf{F1@20} \\
        \midrule
            \multirow{3}{*}{\multirowcell{3}[1.5em]{\text{EntailTree}}} &
            \text{BE}-fix & 20.1 & 15.0 & 9.7 \\
            & \text{BE}-dyn & 24.9 & 19.4 &16.9  \\
            & \text{\ours}-fix & 33.6 & 23.5 & 13.8 \\ 
            & \text{\ours}-dyn & 48.2 & 52.1 & 52.5 \\ 
        \midrule
            \multirow{3}{*}{\multirowcell{3}[1.5em]{\text{StrategyQA}}} &
            \text{BE}-fix & 22.8 & 15.6 & 9.3 \\
            & \text{BE}-dyn & 38.1 & 36.9 & 36.5  \\
            & \text{\ours}-fix & 30.1 & 19.3 & 11.2\\ 
            & \text{\ours}-dyn & 41.9 & 44.3 & 46.6  \\ 
    
        \midrule
            \multirow{3}{*}{\multirowcell{3}[1.5em]{\text{EG-Open}}} &
            \text{BE}-fix & 24.6 & 20.1 & 13.1 \\
            & \text{BE}-dyn &27.0 & 24.6 & 25.4 \\
            & \text{\ours}-fix & 26.9 & 22.1 & 15.1\\ 
            & \text{\ours}-dyn & 35.5 & 40.0 & 41.5 \\ 
        \bottomrule
    \end{tabular}
    \label{table: dynamic}
\end{table}

\subsubsection{Fixed vs. Dynamic Multi-Hop Settings} \label{app: fixed_dynamic_result}
Table~\ref{table: dynamic} shows the retrieval sequence F1 score (F1@k) of Fixed and Dynamic Multi-Hop Settings, where $k$ is the number of maximum retrieval steps, and we evaluate $k$=5, 10, and 20. For all three evaluations and both the bi-encoder approach and \ours, models trained in the dynamic setting show higher scores than those trained in the fixed setting, emphasizing the importance of using dynamic multi-hop retrieval settings to solve multi-hop retrieval tasks near a real-world scenario.

\subsubsection{RT-Open Results}
\label{app: ruletaker_result}
\begin{figure}[h]
\centering
\begin{subfigure}[b]{0.45\textwidth}
\includegraphics[width=\textwidth]{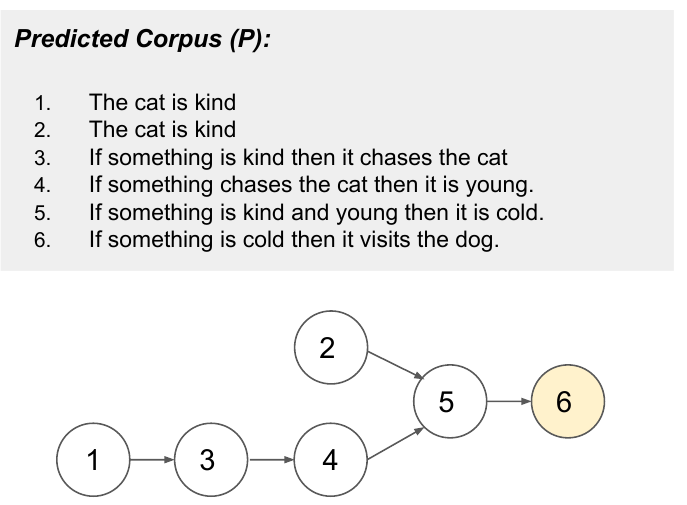}
\end{subfigure}
\begin{subfigure}[b]{0.45\textwidth}
\includegraphics[width=\textwidth]{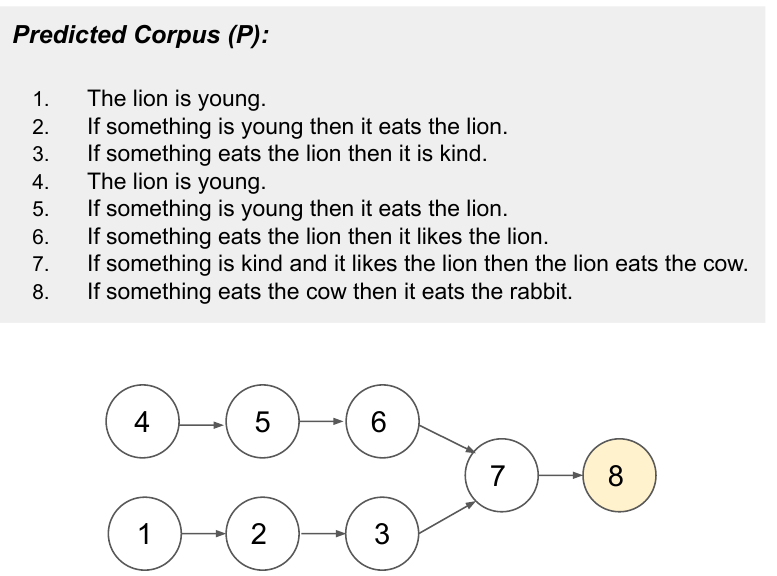}
\end{subfigure}\\
\caption{\fontsize{7.5}{10}\footnotesize Success Examples of \ours}
\label{fig:ar_success}
\end{figure}

\begin{figure}[h]
\centering
\begin{subfigure}[b]{0.45\textwidth}
\includegraphics[width=\textwidth]{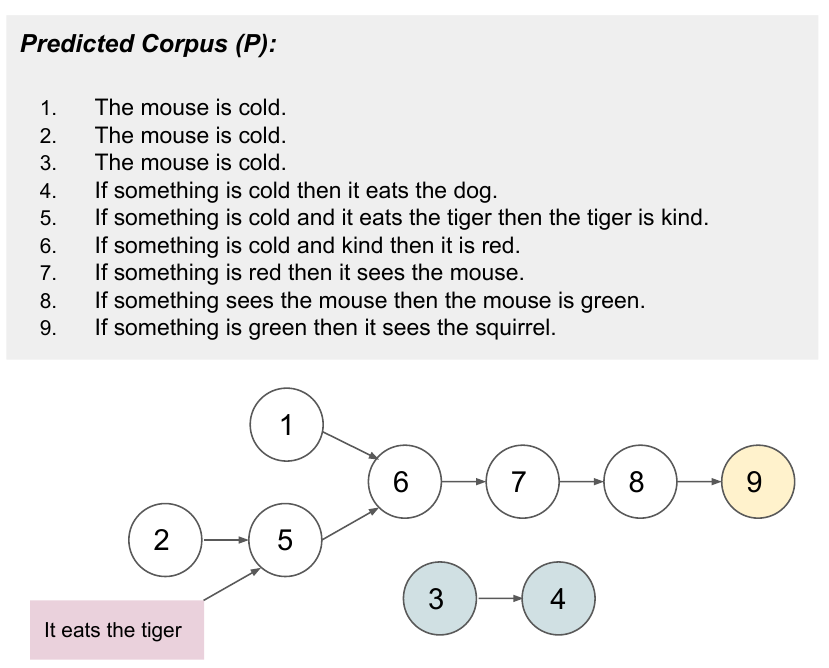}
\end{subfigure}
\begin{subfigure}[b]{0.45\textwidth}
\includegraphics[width=\textwidth]{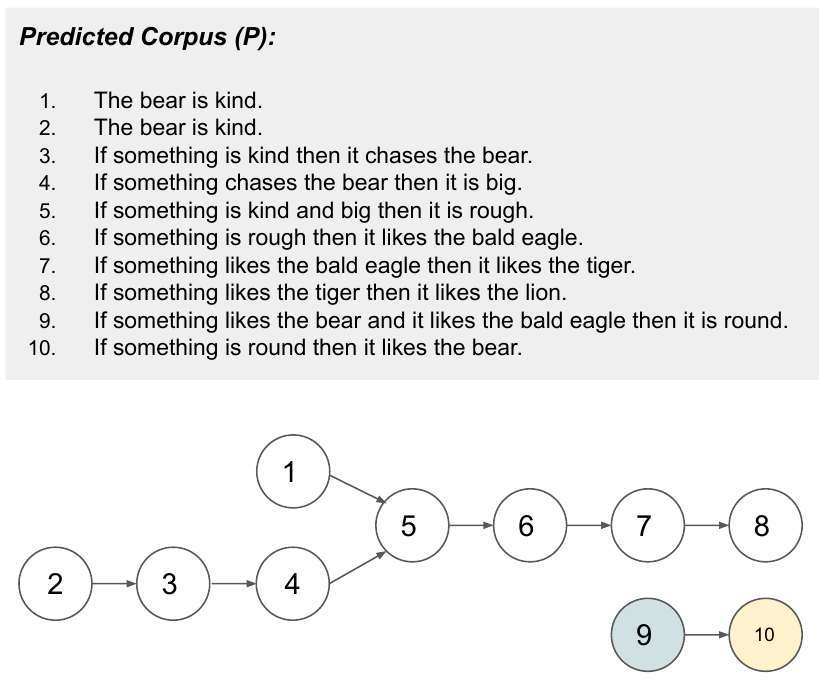}
\end{subfigure}\\
\caption{\fontsize{7.5}{10}\footnotesize Failure Examples of \ours where blue nodes indicate the the leftover nodes and red are the missing nodes}
\label{fig:ar_failure}
\end{figure}

\begin{figure}[!tbp]
\centering
\begin{subfigure}[b]{0.45\textwidth}
\includegraphics[width=\textwidth]{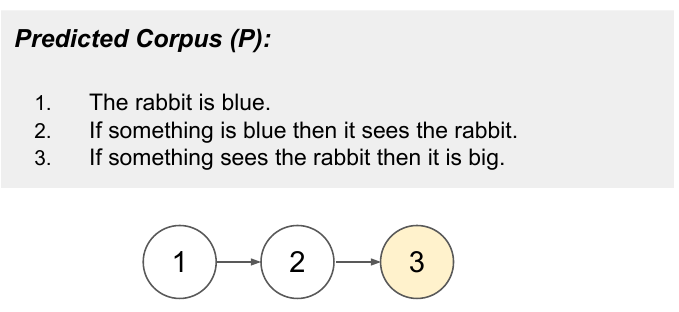}
\end{subfigure}
\begin{subfigure}[b]{0.45\textwidth}
\includegraphics[width=\textwidth]{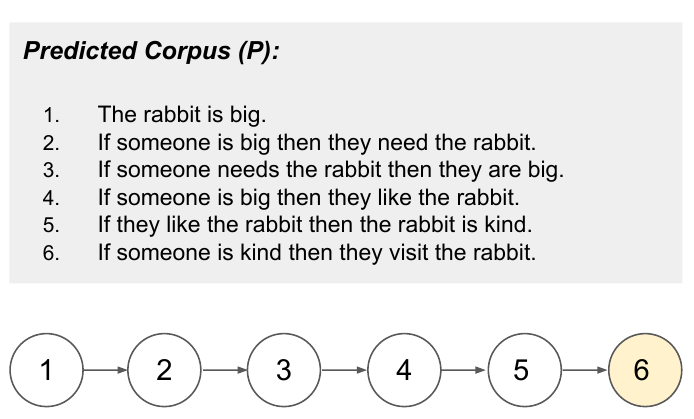}
\end{subfigure}\\
\caption{\fontsize{7.5}{10}\footnotesize Success Examples of Bi-encoder (ST5) Retrieval}
\label{fig:bi_success}
\end{figure}

\begin{figure}[h]
\centering
\begin{subfigure}[b]{0.45\textwidth}
\includegraphics[width=\textwidth]{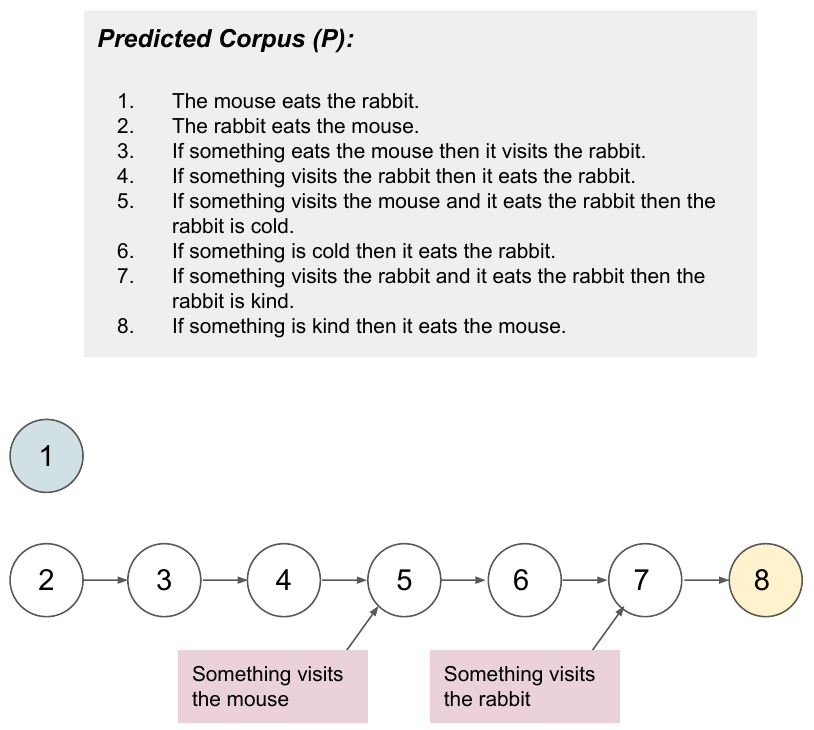}
\end{subfigure}
\begin{subfigure}[b]{0.45\textwidth}
\includegraphics[width=\textwidth]{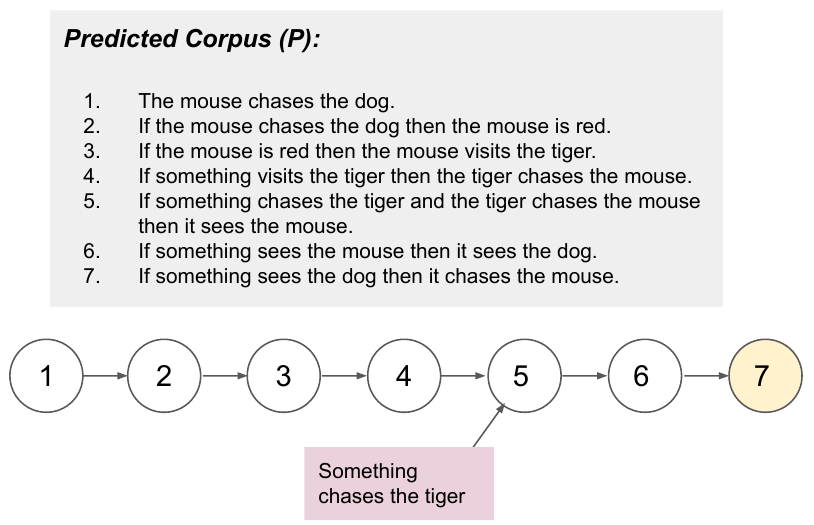}
\end{subfigure}\\
\caption{\fontsize{7.5}{10}\footnotesize Failure Examples of Bi-encoder (ST5) Retrieval where blue nodes indicate the the leftover nodes and red are the missing nodes}
\label{fig:bi_failure}
\end{figure}

The prediction result from the model, predicted corpus ($P$), is in the gray box, and the final node is colored in yellow. The Missing nodes are colored in red, and the leftover nodes are colored in blue. If there is a red or blue node, it means that it failed to construct the reasoning graph. We show two examples for each retrieval method and success and failure cases (missing edge error case) in Figure~\ref{fig:ar_success}, Figure~\ref{fig:ar_failure}, Figure~\ref{fig:bi_success}, and Figure~\ref{fig:bi_failure}.

\subsubsection{Storage Footprint} \label{app: storage}
Table~\ref{table:memory} shows the overall storage footprint of three models: MDR, \ours, and \ours with early stopping. 
Where \ours with early stopping does not generate every word in the retrieval target text but stops generation as soon as the partially generated text can uniquely identify the target text and saves only till the point.
Also, \ours shows higher memory efficiency with a higher decrease rate of the index size (with respect to that of the bi-encoder counterpart) when the granularity of the corpus is small and when items in the given target corpus are similar to one another with the same prefix. HotpotQA, which uses paragraphs as the retrieval unit, has a lower reduction rate of 86.68\%. RT-Open shows the highest decrease rate of 99.9\% since its corpus consists of short texts, and the items in the corpus are highly similar to one another due to its synthetic rule-based data construction process. Shorter retrieval sequences result in a higher index decrease rate since bi-encoder retrievers use a fixed-size dense embedding regardless of the sequence length, whereas \ours stores fewer tokens for shorter sequences.
Moreover, when the number of retrieval sequences in the corpus increases, the storage footprint of the bi-encoder model increases linearly, whereas that of \ours increases more slowly as it needs to store only the additional token ids (integers) that are not in the prefix tree.

\begin{table}[t!]
    \centering
    \fontsize{7.5}{10}\selectfont
    \caption{\fontsize{7.5}{10}\footnotesize Storage footprint for \ours and bi-encoder in GB. \ours* is a model of \ours with early stopping. \ours shows average of 73.49\% reduction on total memory usage compare to the bi-encoder model.}
    \begin{tabular}{cl c c c}
        \toprule
        \textbf{Dataset} & \textbf{Model} &
        \textbf{Retriever} & \textbf{Index} & \textbf{Total} \\
        \midrule
            \multirow{3}{*}{\multirowcell{3}[1.5em]{\text{HotpotQA}}} &
            \text{MDR} & 0.48 & 15.33 & 15.81\\
            & \text{\ours} & 2.75 & 2.04 & 4.79\\
            & \text{\ours*} & 2.75 & 0.20 & 2.95 \\
        \bottomrule
    \end{tabular}
    \label{table:memory}
\end{table}

\subsubsection{DSI} \label{app: DSI}
As DSI~\citep{Tay2022TransformerMA} is not open-sourced, we reproduce both the model and the dataset (NQ-10k) ourselves. In Table~\ref{table: DSI_NQ}, we show results of DSI* and DSI which DSI* is our reproduced model and DSI is the model from the original paper. For the Atomic Docid method, DSI* shows near twice the performance of DSI in both Hits@1 and Hits@10. For the Naive String Docid method, DSI* shows near 80\% of DSI performance in Hits@1 and Hits@10. 

In Table~\ref{table: DSI}, we could see that DSI especially shows a low recall score in StrategyQA dataset, which has a corpus set four times larger than the other two datasets (EG-Open and EntailBank). Such tendency of the performance degradation as the size of the target corpus increases can also be seen in the DSI paper when comparing the result between NQ-10k and NQ-320k. These results suggest the possible difficulty of expanding to a larger corpus set in DSI unlike \ours.

\begin{table}[t!]
    \centering
    \fontsize{7.5}{10}\selectfont
    \caption{\fontsize{7.5}{10}\footnotesize Result of NQ-10k dataset of our reproduced DSI model (DSI*). DSI are results from Table 3 of \citet{Tay2022TransformerMA}. Although we tried to replicate the same setting as in DSI since the dataset NQ-10k is unreleased, the DSI and DSI* datasets may differ. We use T5-base with initial checkpoint from \citet{Wolf2020TransformersSN}. We did not report the score of the Semantic String Docid method since both Hits@1 and Hits@10 of DSI* are very low.} 
    \begin{tabular}{cl c c c}
        \toprule
        \textbf{Method} & \textbf{Model} &
        \textbf{Hits@1} & \textbf{Hits@10} \\
        \midrule
            \multirow{2}{*}{\multirowcell{2}[0.7em]{\text{Atomic Docid}}} &
            \text{DSI} & 13.0 & 38.4 \\
            & \text{DSI*} & 38.2 & 60.1 \\
            \midrule
            \multirow{2}{*}{\multirowcell{2}[0.7em]{\text{Naive String Docid}}} &
            \text{DSI} & 28.1 & 48.0\\
            & \text{DSI*} & 22.6 & 37.0 \\
        \bottomrule
    \end{tabular}
    \label{table: DSI_NQ}
\end{table}

\begin{figure*}[t!]
\centering
\includegraphics[width=0.75\linewidth]{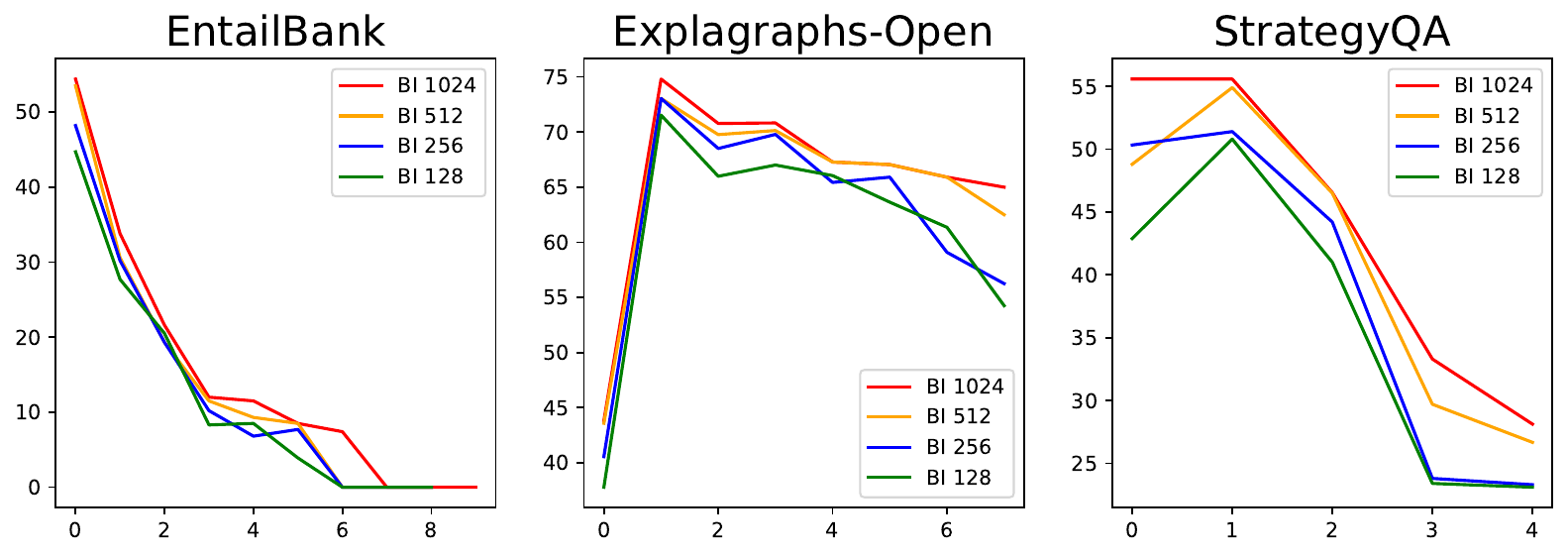}
\caption{\fontsize{7.5}{10}\footnotesize We plot hop-R@5 (y-axis) over number of hops in multi-hop retrieval task (x-axis) in the figure. As we experiment on ST5 using T5-large, the initial embedding size is 1024. We experiment by reducing or retaining the 1024 embedding to 1024 (red), 512 (orange), 256 (blue), and 128 (green) dimensions by adding a linear layer at the end of the model. For all three datasets, we can see that performance of the bi-encoder tends to degrade as (1) the number of hops increases after a certain threshold value and (2) as the size of embedding decreases.}
\label{fig: bi_emb_size}
\end{figure*}

\subsection{Analysis}

\subsubsection{Manual Analysis on HotpotQA}  \label{app: hotpotqa}
We conduct manual analysis on HotpotQA by comparing the top-2 prediction result of the \ours and MDR, a bi-encoder retrieval model. From the two question categories in HotpotQA (bridge and comparison questions), we manually inspect 30 sampled examples where one model fits and the other is wrong. MDR mostly got wrong by missing the second hop item though it got the first hop correct and \ours was wrong for cases where the first-hop item is not written explicitly in the query but by sharing a specific part of a sentence. When the item is written explicitly in the query, \ours tend to get it correct, which shares with the result that \ours shows a higher score on comparison questions than MDR. We suggest this result is because \ours can directly cross-encode between the input and the output without any information loss.

To be specific, we divide the error case into four:\\
(1) When the first-hop retrieval item is not written explicitly in the query but by sharing a specific part of a sentence. \\
(2) Though it is written explicitly in the query, it retrieves the wrong document by giving attention to an irrelevant part of the query. \\
(3) Detail of the title is wrong (i.e., when the gold document has the title \textit{Do you Love Me (Not That I Can Dance)}, the model retrieves a document with the title \textit{Do you Love Me (2NE1 song)} instead; when \textit{do you love me} is in a query, the model misses to understand the details correctly.)\\
(4) The retriever got the first hop correct but failed to retrieve the second hop item correctly. \\
When comparing the number of models matched in the bridge question with each error case, among the four cases, MDR is often wrong in the second (1.3 times) and fourth cases (2.2 times), and the \ours is most often wrong in the first case (6 times) along with the third case (2.8 times)\footnote{the value in parentheses shows the ratio of the error rate compared to the other model}.

\subsubsection{BottleNeck Problem in Bi-Encoder Models} \label{app: emb_size}
Previous work has shown the inherent limitations of bi-encoder approaches; by encoding all information in given text into fixed-size embedding, it has shown a bottleneck problem~\citep{luan2020sparse}. 
By adding a linear layer at the top of the model and decreasing the dimension, we could see that such a bottleneck problem still holds in our bi-encoder models in multi-hop retrieval tasks. As in Figure~\ref{fig: bi_emb_size}, as the embedding size decreases from 1028 to 128, hop-R@5 of the bi-encoder retriever monotonically decreases. The x-axis is the number of hops in multi-hop retrieval tasks and the y-axis is the score of hop-R@5. For all three datasets (EntailBank, EG-Open, and StrategyQA), we can see that performance of the bi-encoder tends to degrade as (1) the number of hops increases after a certain threshold value and (2) as the size of embedding decreases.

\end{document}